\documentclass[useAMS,usenatbib]{mn2e}
\usepackage[fleqn]{amsmath}
\usepackage{graphicx}
\usepackage{color}

\title[Bright Compact bulges]{Bright Compact Bulges (BCBs) at intermediate redshifts}
\author[Sonali Sachdeva and Kanak Saha]{Sonali Sachdeva$^{1}$, Kanak Saha$^{1}$\\
$^{1}$Inter-University Centre for Astronomy and Astrophysics, Pune 411007, India}
\begin{document}

\date{in original form 2017 September 17}

\pagerange{\pageref{firstpage}--\pageref{lastpage}} \pubyear{2014}

\maketitle

\label{firstpage}

\begin{abstract}

Studying bright ($M_B<-20$), intermediate-redshift ($0.4<z<1.0$), disc dominated ($n_B<2.5$) galaxies from HST/ACS and WFC3 in Chandra Deep Field South, in rest-frame {\it B} and {\it I}-band, we found a new class of bulges which is brighter and more compact than ellipticals. 

We refer to them as ``Bright, Compact Bulges" (BCBs) - they resemble neither classical nor pseudo-bulges and constitute $\sim12$\% of the total bulge population at these redshifts. Examining free-bulge $+$ disc decomposition sample and elliptical galaxy sample from Simard et al. (2011), we find that only $\sim0.2$\% of the bulges can be classified as BCBs in the local Universe.

Bulge to total ratio $(B/T)$ of disc galaxies with BCBs is (at $\sim0.4$) a factor of $\sim2$ and $\sim4$ larger than for those with classical and pseudo bulges. BCBs are $\sim2.5$ and $\sim6$ times more massive than classical and pseudo bulges. Although disc galaxies with BCBs host the most massive and dominant bulge type, their specific star formation rate is 1.5-2 times higher than other disc galaxies. This is contrary to the expectations that a massive compact bulge would lead to lower star formation rates.

We speculate that our BCB host disc galaxies are descendant of massive, compact and passive elliptical galaxies observed at higher redshifts. Those high redshift ellipticals lack local counterparts and possibly evolved by acquiring a compact disc around them. The overall properties of BCBs supports a picture of galaxy assembly in which younger discs are being accreted around massive pre-existing spheroids.

\end{abstract}


%%%%%%%%%%%%%%%%%%%%%%%%%%%%%%%%%%%%%%%%%%%%%%%%%%%%%%%%%%%%%%%%%%%%%%%%%%%%%%%%%%%%%%%%%%%%%%%%%%%%%%%%%%%%%%%%%%%%%%%%%%%%%%%%%%%%%%%%%%%%%%%

\section{Introduction}
\label{sec:intro}

Bulges in disc galaxies, earlier known to be similar to ellipticals \citep{DaviesandIllingworth1983,Renzini1999}, were later discovered to be present with disc like properties as well \citep{Kormendy1993,AndredakisandSanders1994,KormendyandKennicutt2004}. Since then, bulges are understood to be classifiable into either being {\it classical}, i.e., having properties closer to ellipticals, or {\it pseudo}, i.e., having properties closer to disks. The classification criteria are thus based on probing the similarity of their properties (mainly surface brightness profile, scaling relations, internal structure, kinematics, stellar population) with those of ellipticals and discs \citep{Fisher2006,FisherandDrory2008,FisherandDrory2016,KormendyandBender2012,Kormendy2016}.

Based on these criteria, bulge classification has been carried out on an extensive scale at local redshifts \citep{Gadotti2009,Simardetal2011,LacknerandGunn2012,Meertetal2015}. For high redshifts, i.e., $z>1$, lately there have been a handful of attempts at decomposing the galaxies into bulge and disc component \citep{Bruceetal2014,Langetal2014,Margalef-Bentaboletal2016}, however, bulges have not been further classified mainly due to lack of resolution and the fact that galaxies begin to resemble the Hubble sequence only closer to $z\sim1$ \citep{Buitragoetal2013,Mortlocketal2013,Huertas-Companyetal2016,Margalef-Bentaboletal2016}.

For $z<1$, galaxies can be imaged in the optical at similar resolution, i.e, $1$ kpc in $\sim 3-4$ pixels, as that achieved by SDSS for local galaxies. This is also the period over which galaxies gained more than half of their present stellar mass and size \citep{Trujilloetal2011,Marchesinietal2014,Ownsworthetal2014}; and underwent morphological transformations, predominantly bulge growth, which aided them to reach stable structures observed at the present epoch \citep{Oeschetal2010,Tascaetal2014,Sachdevaetal2017}. 

In \citet{Sachdevaetal2017} (SS-17, hereafter) we examined the properties of bulges and their host discs, at intermediate ($0.4<z<1.0$) and lower ($0.02<z<0.05$) redshifts, in rest-frame optical and infrared wavelengths. Classifying the bulges into classical and pseudo, according to the Kormendy relation, we presented their evolution from $z\sim1$ to $z\sim0$. In the course of that work, we found a class of bulges which are brighter and more compact that elliptical galaxies. We refer to these ``Bright, Compact Bulges" as BCBs, as they are neither classical bulges (i.e. similar to elliptical galaxies) nor pseudo-bulges (i.e. dimmer and less denser than elliptical galaxies). In this work, we report our study of this new class of bulges, i.e., BCBs, which may provide significant insight regarding the processes of galaxy formation at intermediate redshifts. 

The paper is divided as follows: in Section 2 we present the identification of the new class of bulges (BCBs), illustrate the robustness of our measurements and describe the selection of a local sample for comparison; in Section 3.1 we illustrate the distinctness of BCBs with respect to other bulges; in Section 3.2 we present evidence of their near-absence in the local Universe; in Section 4 we summarize the findings and discuss various possibilities relating to their formation and evolution. We consider a flat $\Lambda$-dominated Universe with $\Omega_{\Lambda}=0.73$, $\Omega_m=0.27$ and $H_o=71$ km sec$^{-1}$ Mpc$^{-1}$. All magnitudes are in AB system.

%%%%%%%%%%%%%%%%%%%%%%%%%%%%%%%%%%%%%%%%%%%%%%%%%%%%%%%%%%%%%%%%%%%%%%%%%%%%%%%%%%%%%%%%%%%%%%%%%%%%%%%%%%%%%%%%%%%%%%%%%%%%%%%%%%%%%%%%%%%%%%%%%%%%%%

\begin{figure*}
\mbox{\includegraphics[width=65mm]{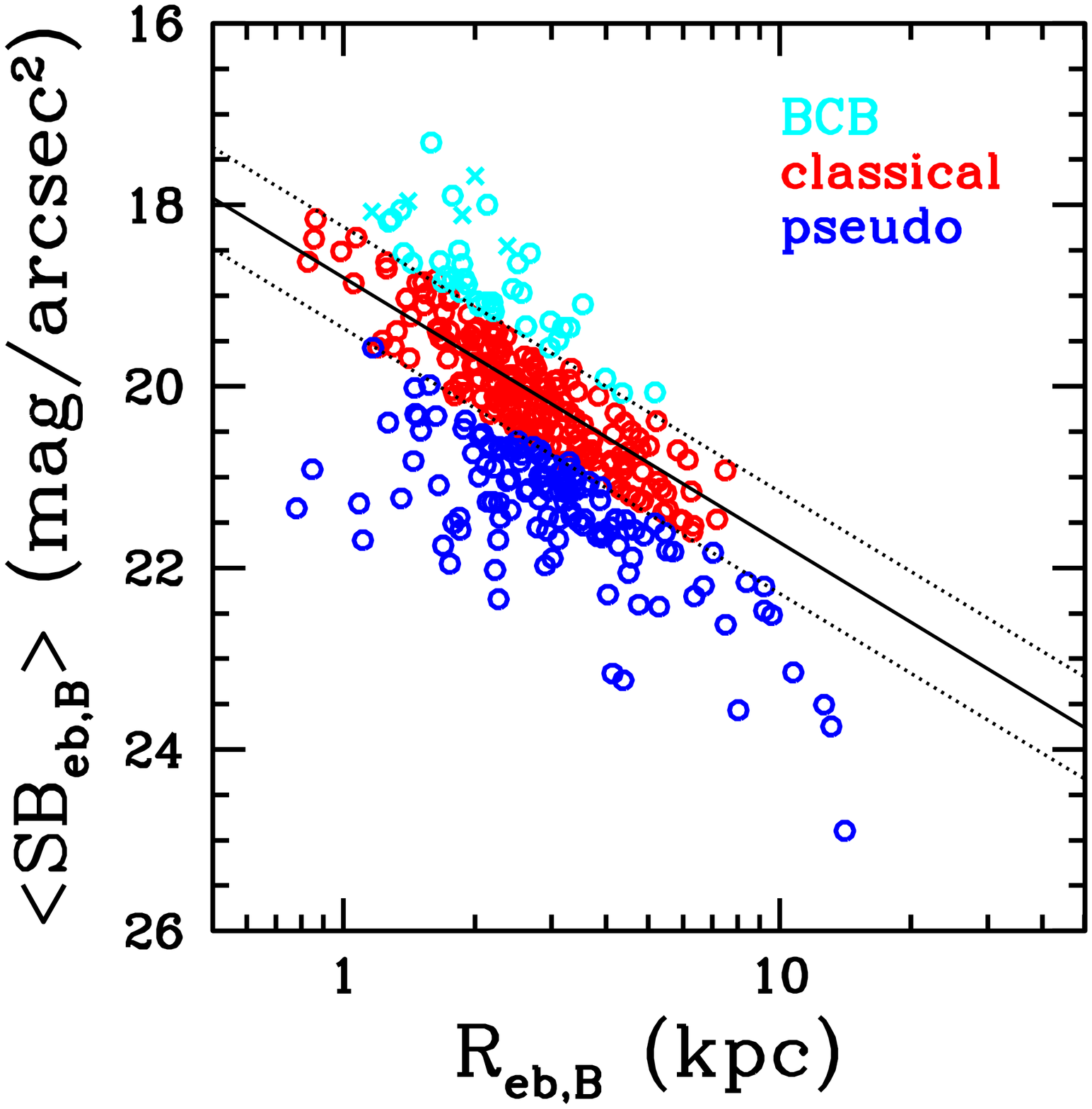}}
\mbox{\includegraphics[width=65mm]{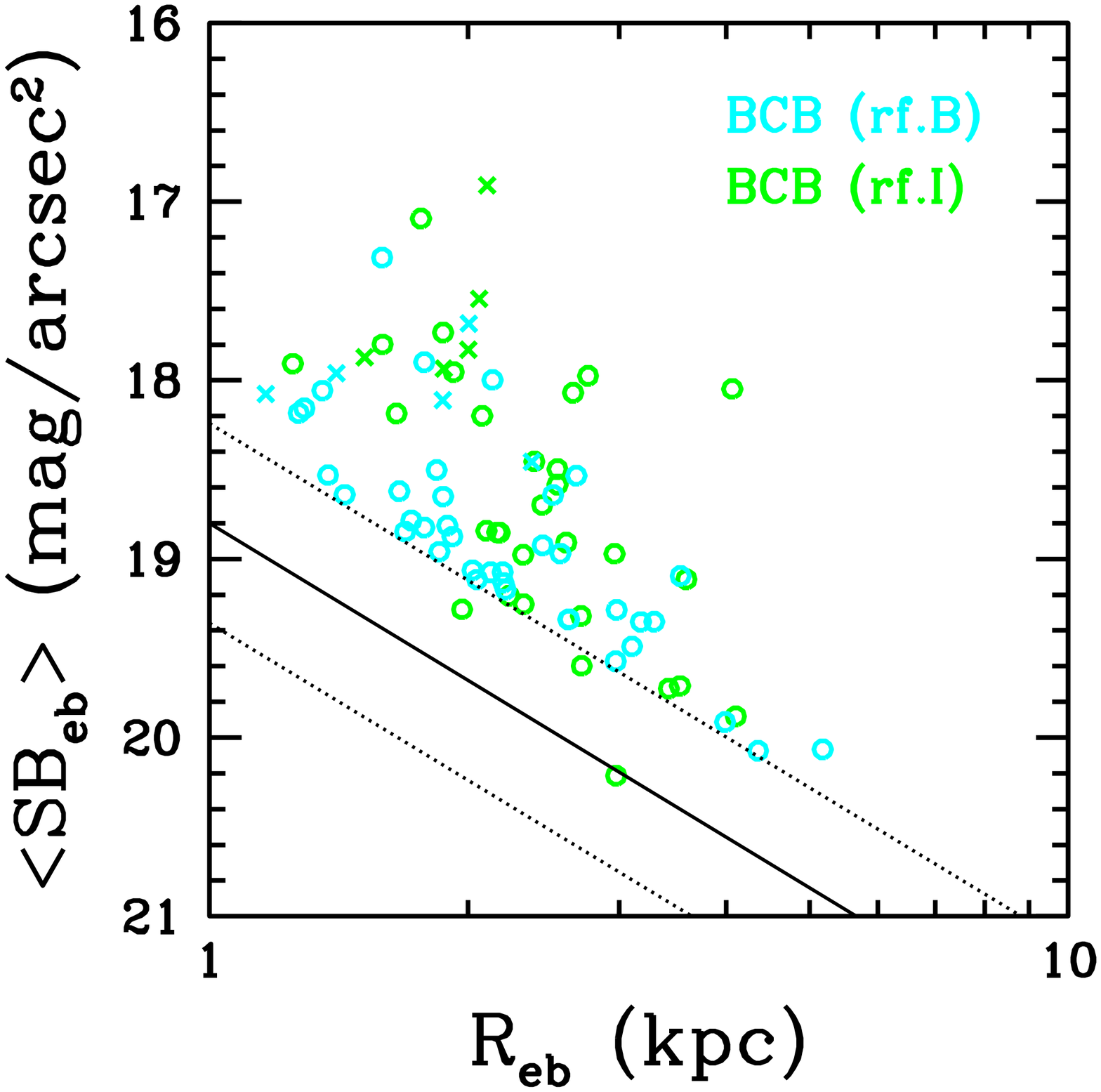}}
\caption{{\it Left panel:} The three bulge types are placed according to their parameters on the Kormendy-plane. The solid line marks the relation followed by elliptical galaxies at the same redshift range (0.4-1.0) and dotted lines mark the three sigma boundaries. Bulges outlier to the upper boundary, i.e, more bright and compact than ellipticals are defined to be BCBs. Five BCBs which were most bright and compact, in both r.f. {\it B} and {\it I} band, are selected for increased focus and marked in the plot as crosses. {\it Right panel:} BCBs are placed in the same plot according to their parameters in both rest-frame (rf.) {\it B} and {\it I}-band.} 
\label{kormendy-selection}
\end{figure*}

\section{Data}
\label{sec:data}

\subsection{Identification of discs with BCBs}

In SS-17, we studied the radial intensity profile of 469 bright, disc dominated, intermediate redshift ($0.4<z<1.0$) galaxies in rest-frame {\it B} and {\it I} band. Rest-frame {\it B}-band images were obtained from GOODS survey, using HST-ACS\footnote{Based on observations obtained with the NASA/ESA HST, which is operated by the Association of Universities for Research in Astronomy, Inc.(AURA) under NASA contract NAS5-26555.} V(F606W), i(F775W) and z(F850LP) filters in Chandra Deep Field South (CDFS) \citep{Giavaliscoetal2004}. Corresponding, rest-frame {\it I}-band images were obtained from CANDELS and 3DHST Treasury Program\footnote{Based on observations taken by the 3D-HST Treasury Program (GO 12177 and 12328) with the NASA/ESA HST} using HST WFC3 J(F125W), JH(F140W) and H(F160W) filters \citep{Groginetal2011,Koekemoeretal2011,Skeltonetal2014}. Selection of the sample according to various criteria, finalization of the images and the procedure for obtaining the radial intensity profile is detailed in that work. 

Examining radial intensity profile of each galaxy, individually, we decomposed it into two components following an intuitive approach. Briefly, first the exponential function (or disc function) was fitted to the exponentially falling part of the intensity profile to obtain disc parameters. Then keeping disc parameters fixed, total intensity function (i.e, sum of the exponential function and S\'ersic function) was fitted to the full profile. This ensured that S\'ersic function fitted only the extra intensity at the centre which was in excess to the underlying disc intensity. The total intensity function, i.e, sum of exponential function (which provides disc parameters) and S\'ersic function (which provides bulge parameters), is given below:

\begin{equation} 
I(r)=I_o\exp(-\frac{r}{r_d})+I_e\exp[-b((\frac{r}{r_e})^{1/n_b}-1)],
\end{equation}

where $I_o$ is the central intensity of the disc, $r_d$ is the scale length of the disc, $r_e$ is the effective radius of the bulge, $I_e$ is the intensity of the bulge at it's effective radius and $n_b$ is the S\'ersic index of the bulge. These bulge and disc parameters, obtained in both rest-frame {\it B} and {\it I}-band for each galaxy, were converted to their intrinsic/absolute values, i.e., radii in kpc, absolute magnitudes and intrinsic surface brightnesses. These conversions \citep[equations involved are illustrated in][]{GrahamandDriver2005,Sachdeva2013} are according to their redshift, cosmology adopted, K-corrections and magnitude zero-points.

Obtaining their intrinsic parameters, we attempted to classify the bulges according to their properties. A bulge is defined to be classical if it's properties are more similar to those of elliptical galaxies and is defined to be pseudo if they are more similar to those of disc galaxies \citep{KormendyandKennicutt2004,Sellwood2014,FisherandDrory2016}. A fundamental property of elliptical galaxies is that their effective radius and surface brightness within that radius follow a tight correlation, at all redshifts, which is known as the Kormendy relation \citep{Kormendy1977}. Thus, classical bulges can be identified as those which lie within the 3-sigma boundaries of Kormendy relation of ellipticals \citep{FisherandDrory2008,KormendyandBender2012}. On the other hand, pseudo bulges are identified as low surface brightness outliers, i.e., those which lie below the 3-sigma boundaries of Kormendy relation for ellipticals \citep{Carolloetal2001,Gadotti2009,Sachdevaetal2015,FisherandDrory2010}.

We followed this classification method in SS-17. The Kormendy relation for ellipticals, at $0.4<z<1.0$, was found, for rest-frame {\it B}-band, using galaxies with global S\'ersic index more than 3.5. Then we plotted our disc dominated sample according to their bulge parameters, in rest-frame {\it B}-band, onto that relation. Note that disc galaxies without bulges, and thus, lacking bulge parameters, did not get included in this exercise \citep[these have been explored in][]{SachdevaandSaha2016}. For our final sample of 358 galaxies, with both bulge and disc parameters, we found that 53\% (189 out of 358) of the bulges were lying within the 3-sigma boundaries, 35\% (126 out of 358) were lying below the 3-sigma boundaries, and interestingly, remaining 12\% (43 out of 358) of the bulges were lying above the 3-sigma boundaries, as shown in Fig.~\ref{kormendy-selection}. Thus, 43 bulges were discovered, at $0.4<z<1.0$, to be brighter and more compact than elliptical galaxies. We refer to these bulges as ``Bright Compact Bulges" (BCBs) throughout this work. BCBs were not probed further in SS-17 as the mandate of that work was to study the growth of classical and pseudo bulges relative to their host discs. In this work, we will explore the properties of BCBs in comparison to other bulges, check for their counterparts in the local Universe and progenitors at high redshifts, to understand the context of their formation and later evolution. 

%%%%%%%%%%%%%%%%%%%%%%%%%%%%%%%%%%%%%%%%%%%%%%%%%%%%%%%%%%%%%

\begin{figure*}
\mbox{\includegraphics[width=50mm]{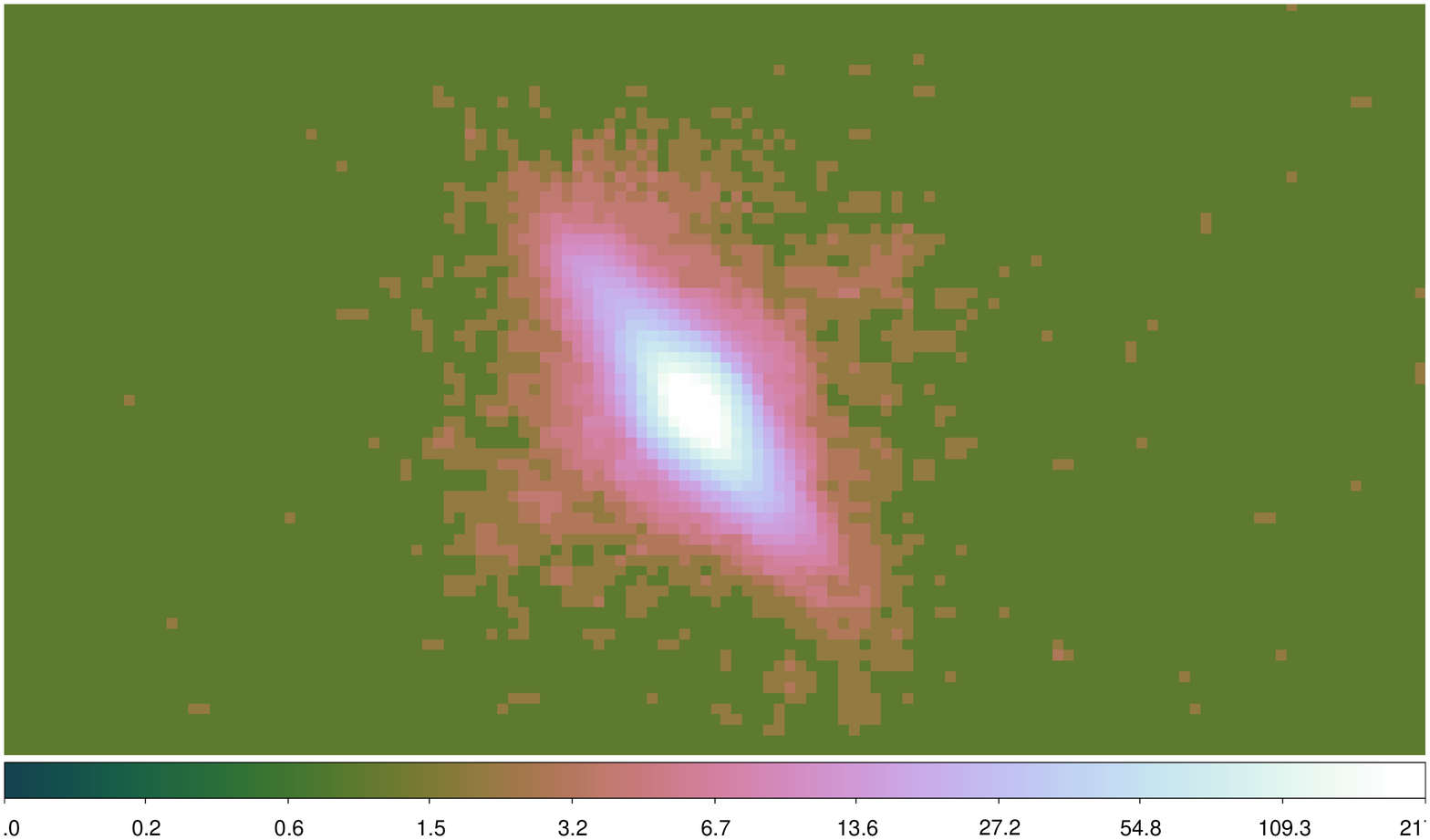}}
\mbox{\includegraphics[width=40mm]{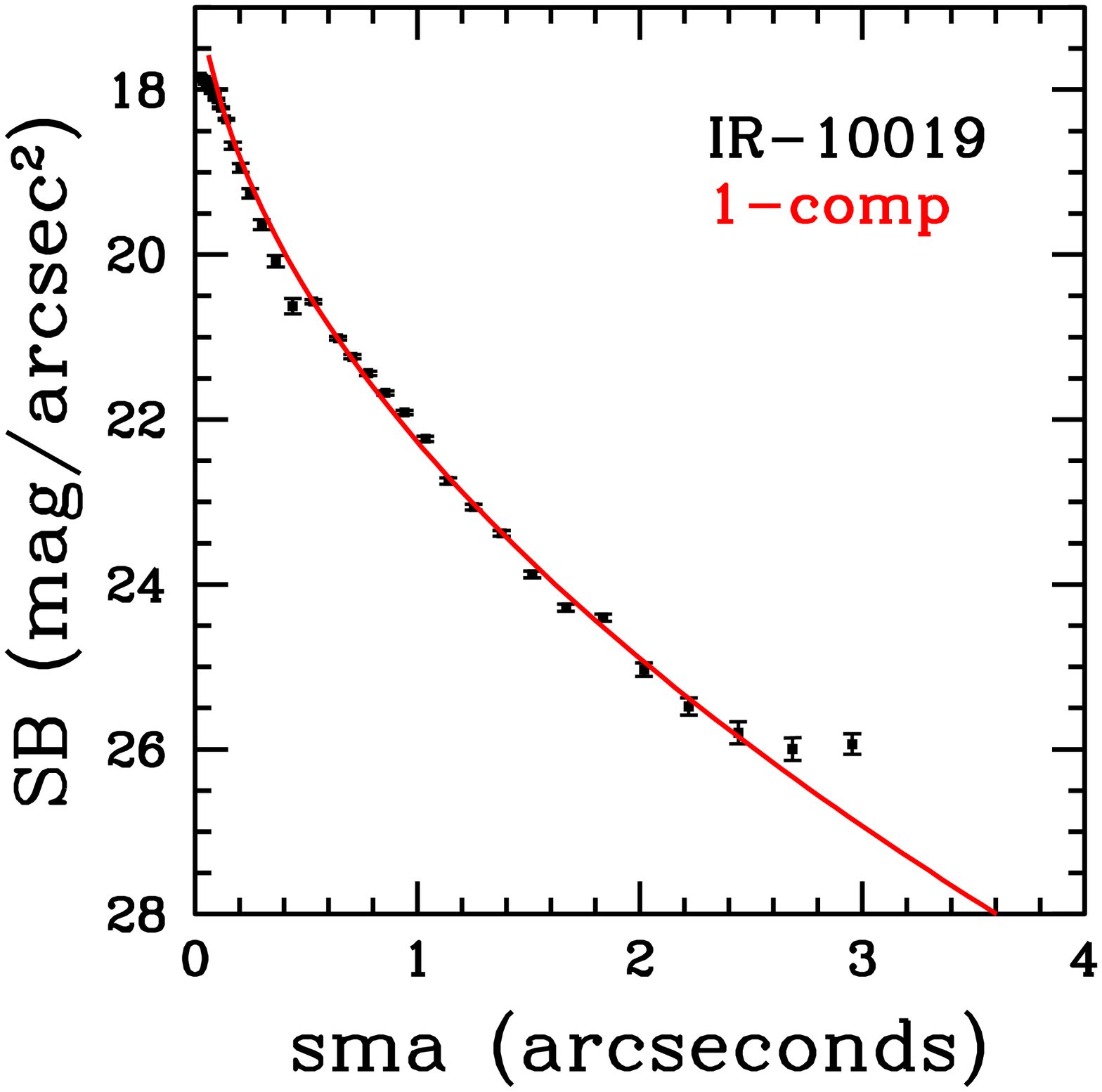}}
\mbox{\includegraphics[width=40mm]{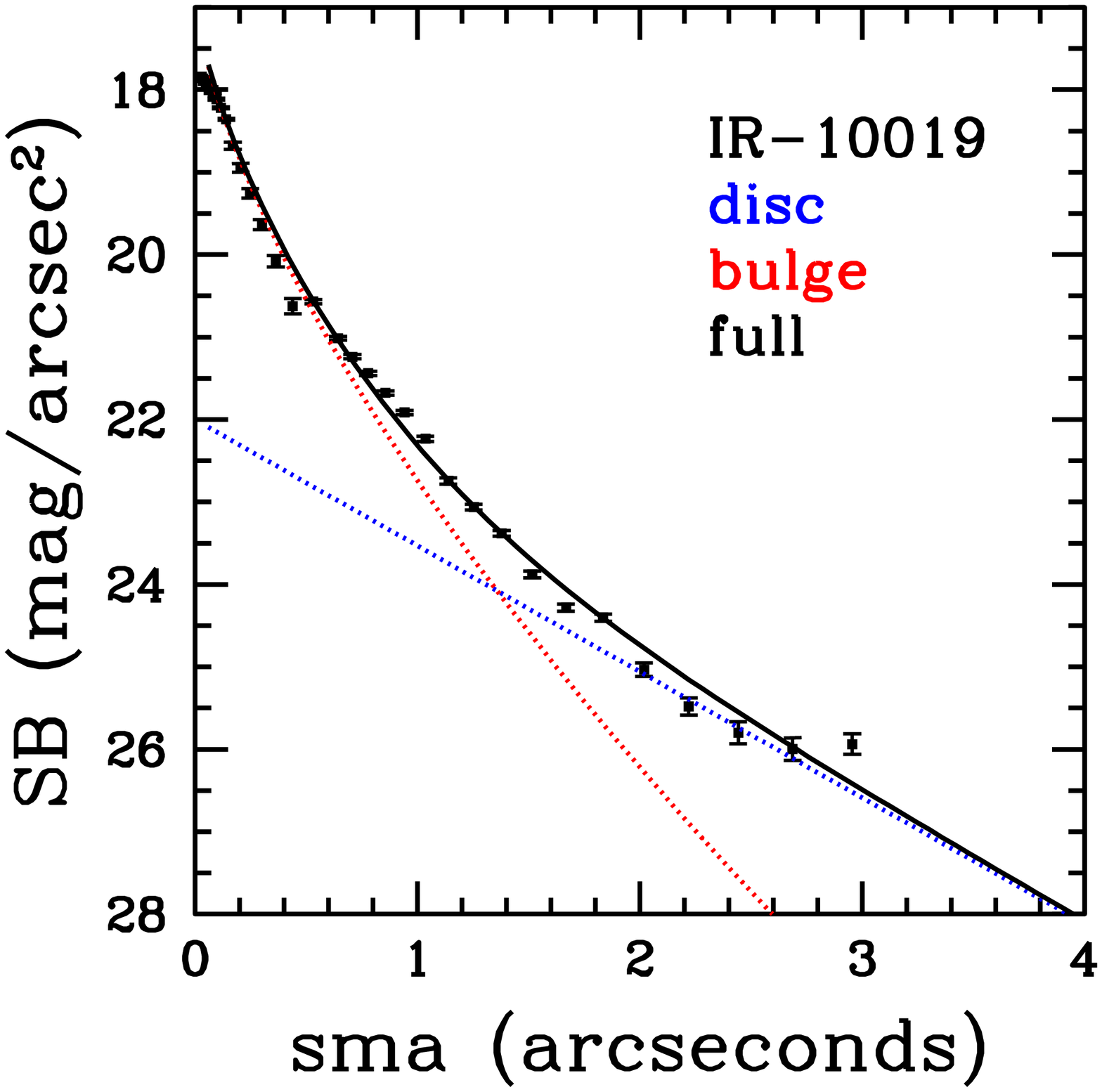}}\\
\mbox{\includegraphics[width=50mm]{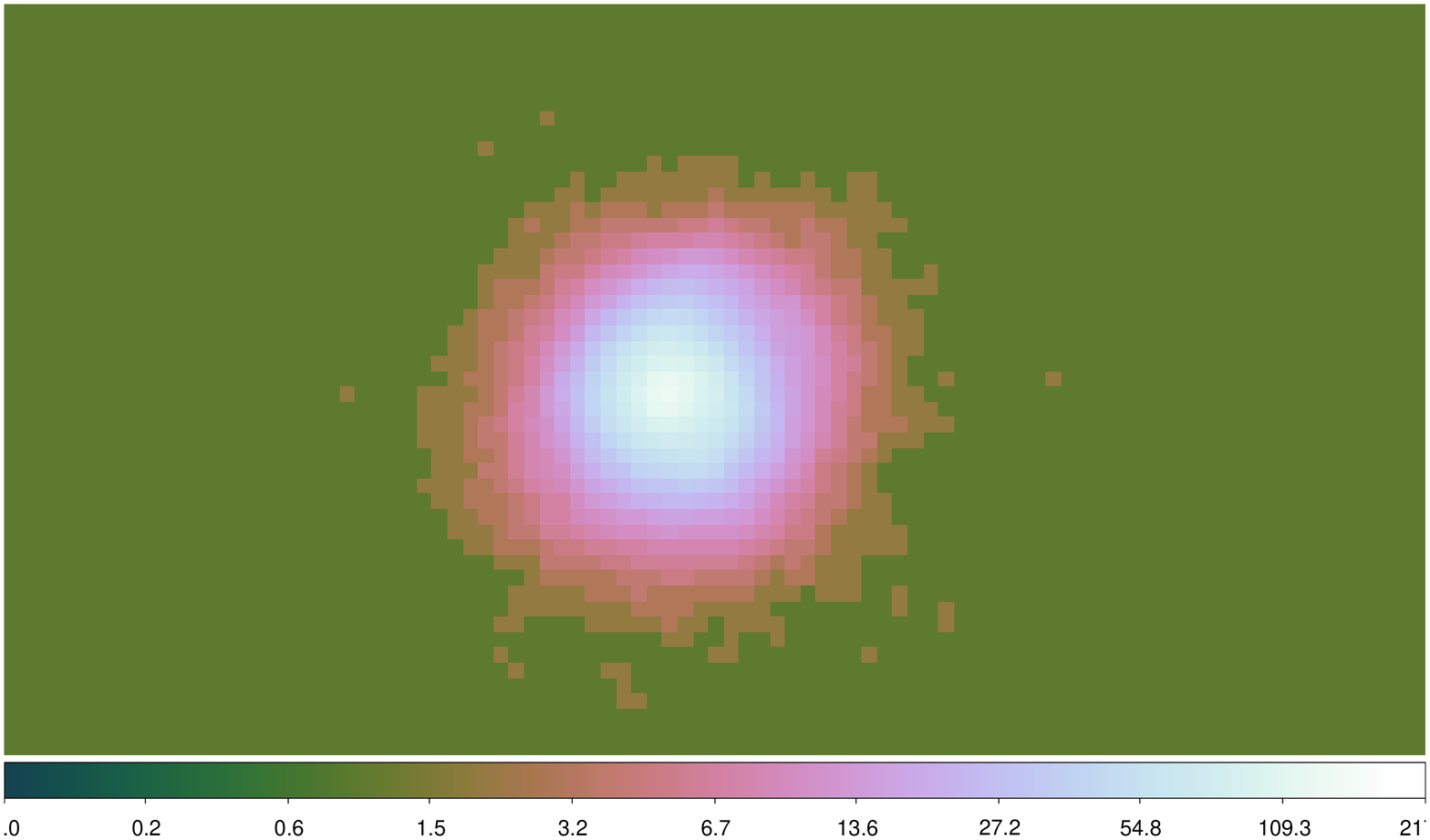}}
\mbox{\includegraphics[width=40mm]{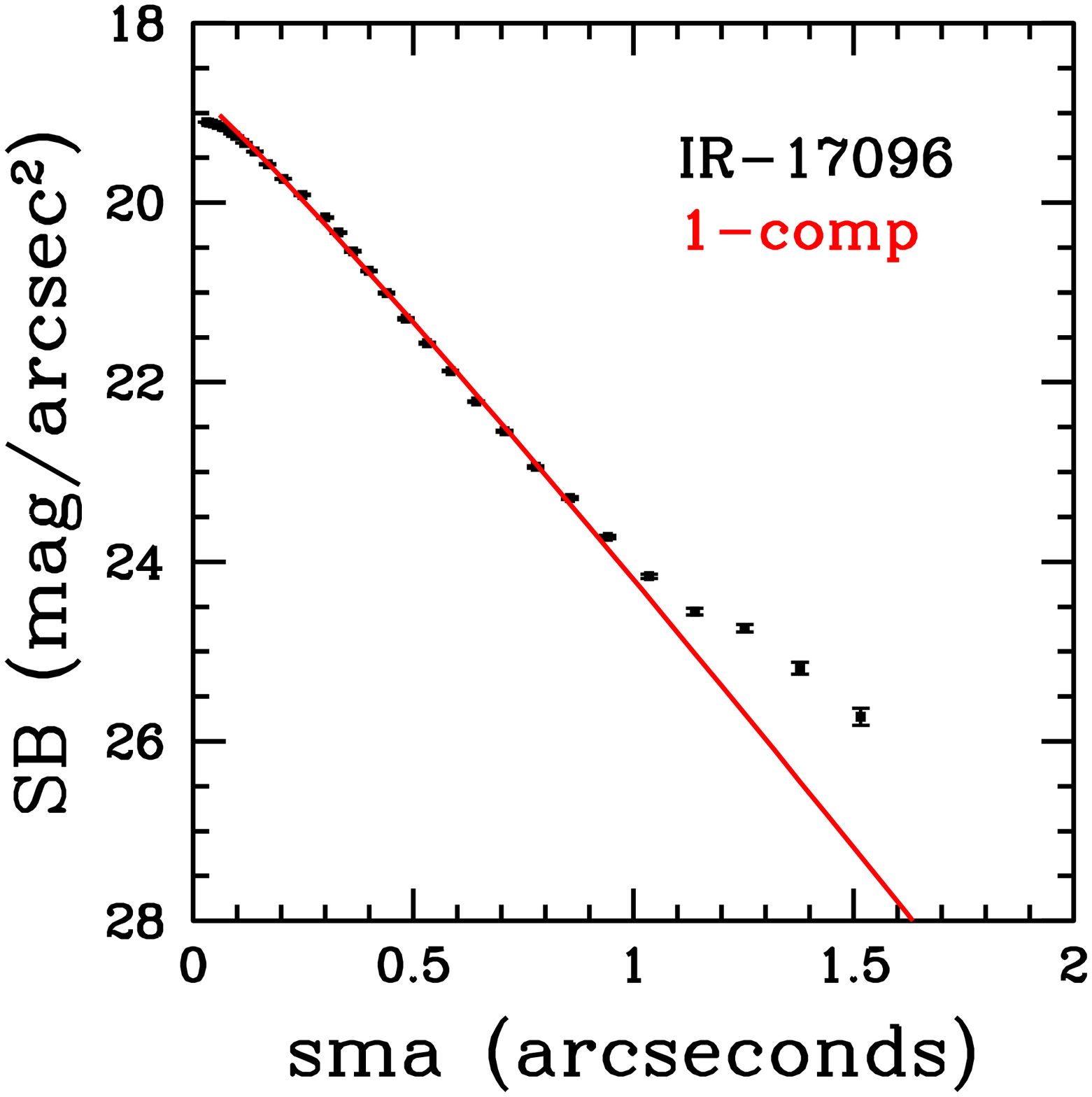}}
\mbox{\includegraphics[width=40mm]{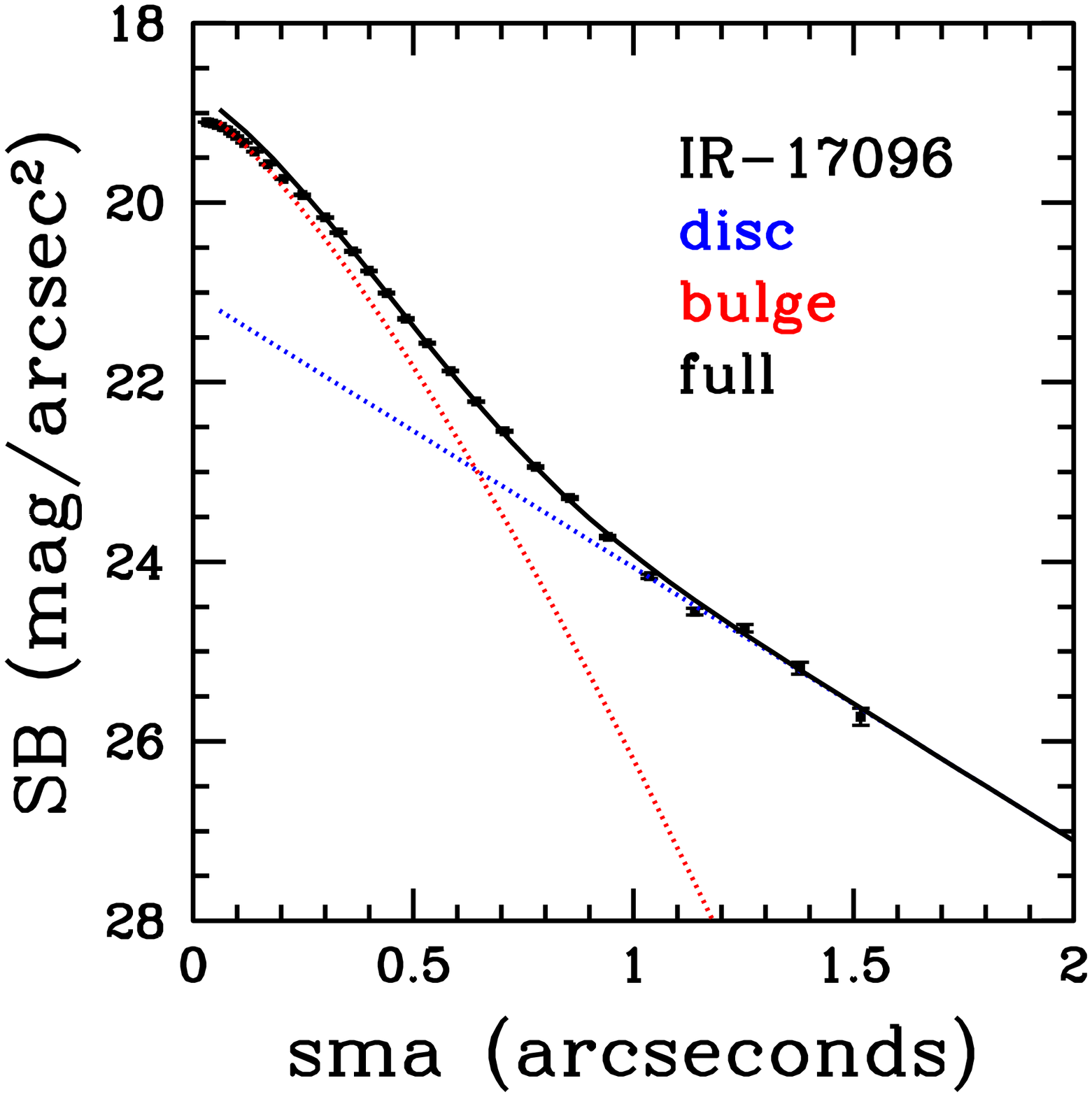}}\\
\mbox{\includegraphics[width=50mm]{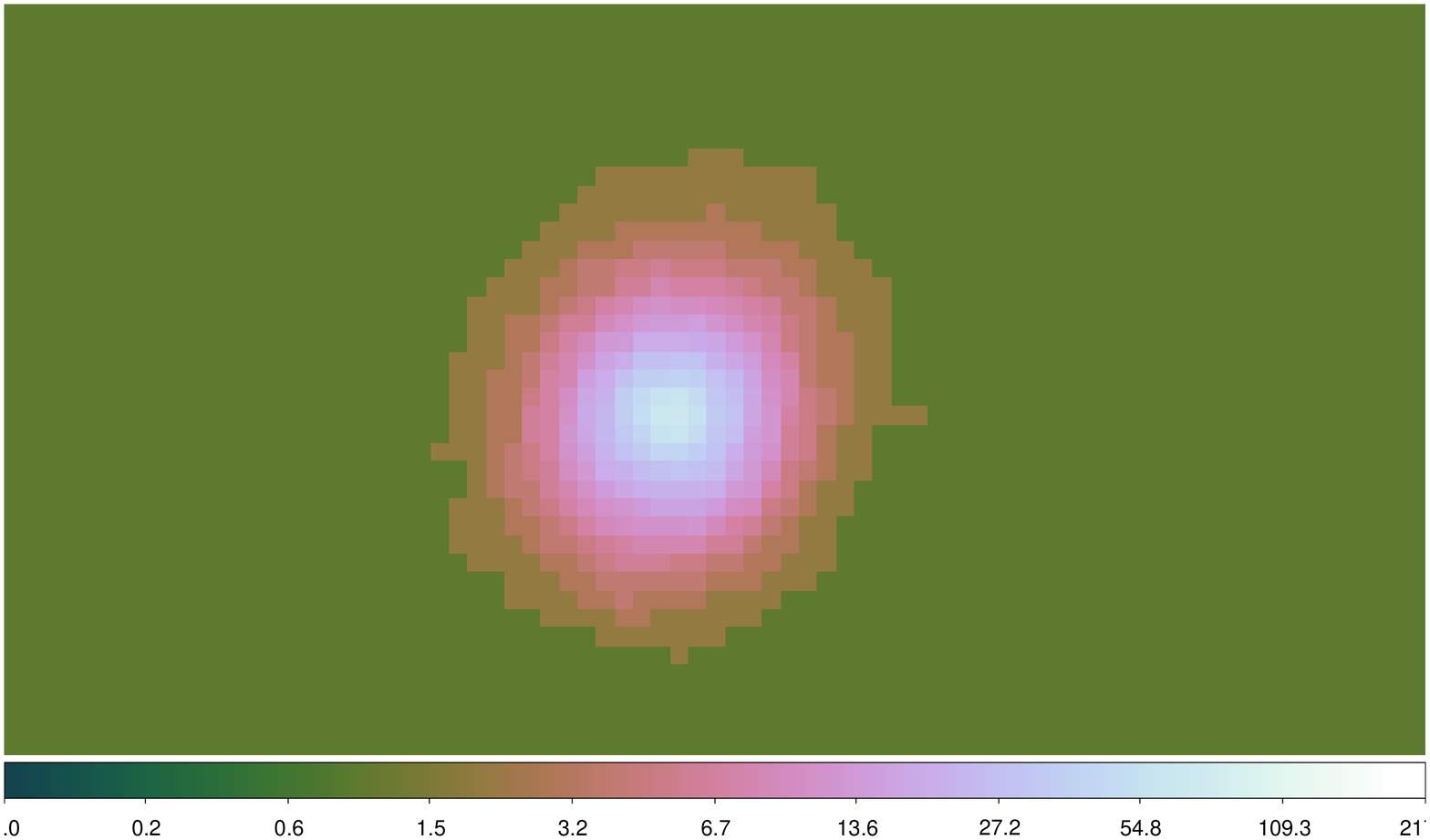}}
\mbox{\includegraphics[width=40mm]{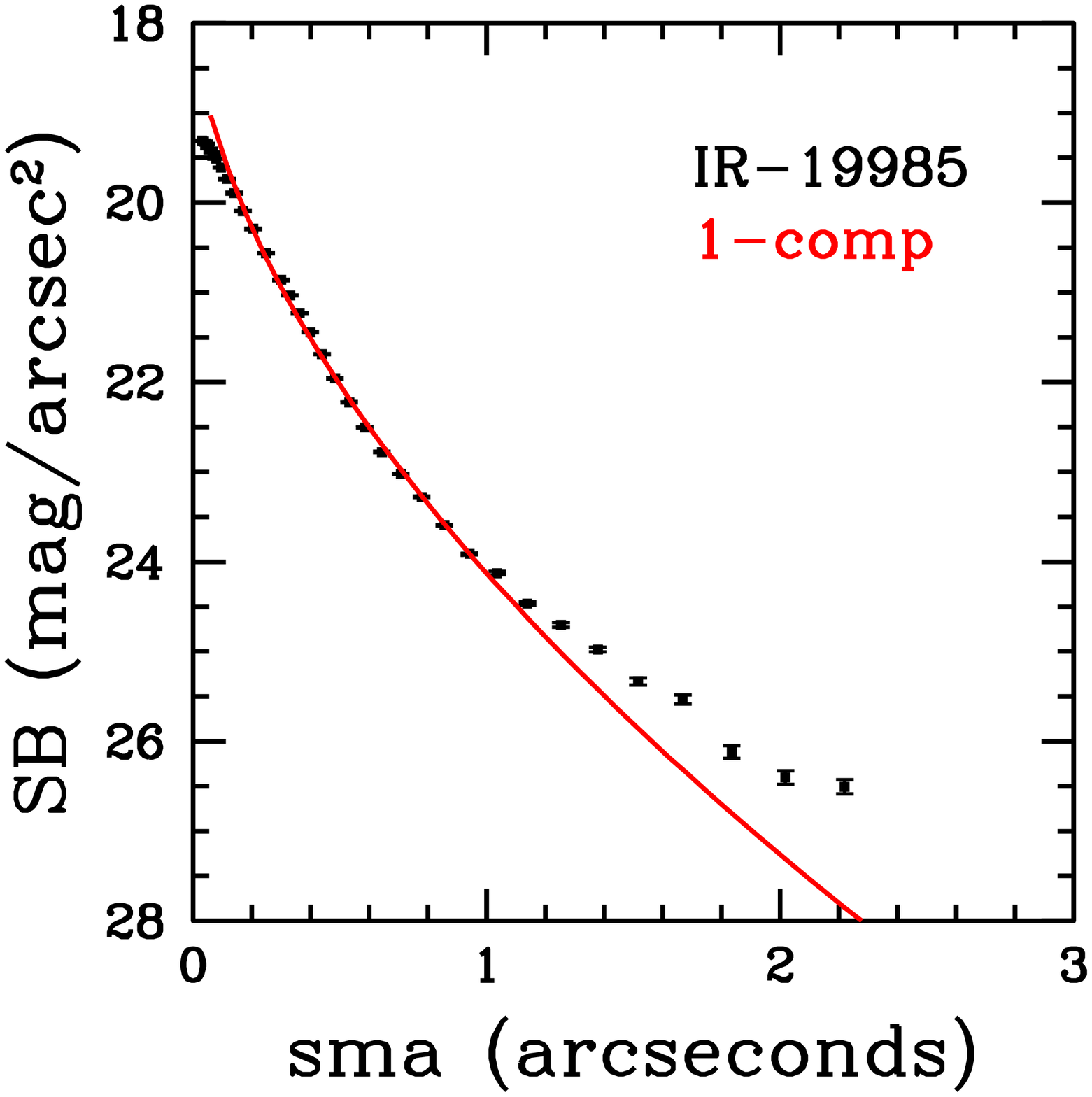}}
\mbox{\includegraphics[width=40mm]{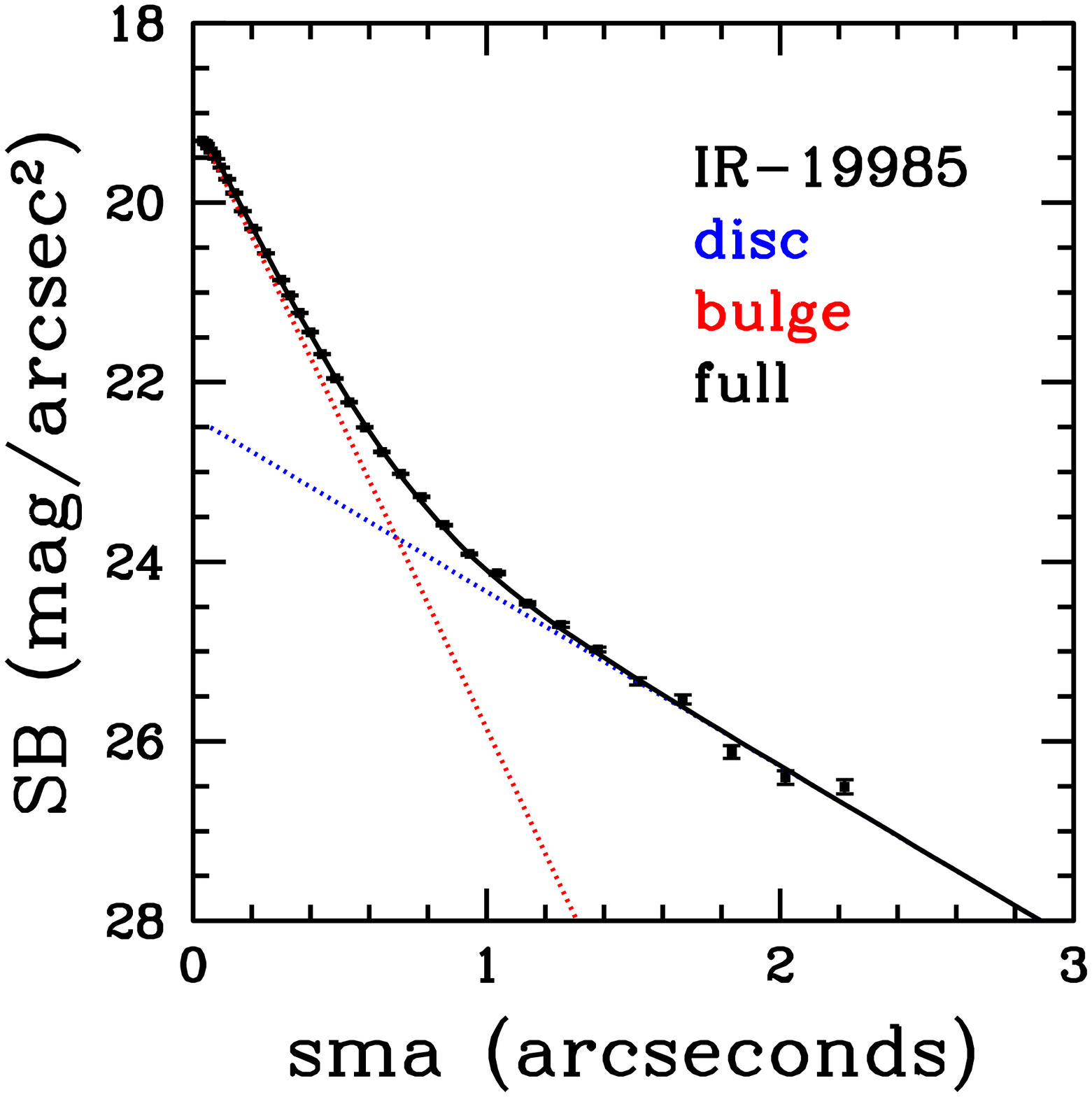}}\\
\mbox{\includegraphics[width=50mm]{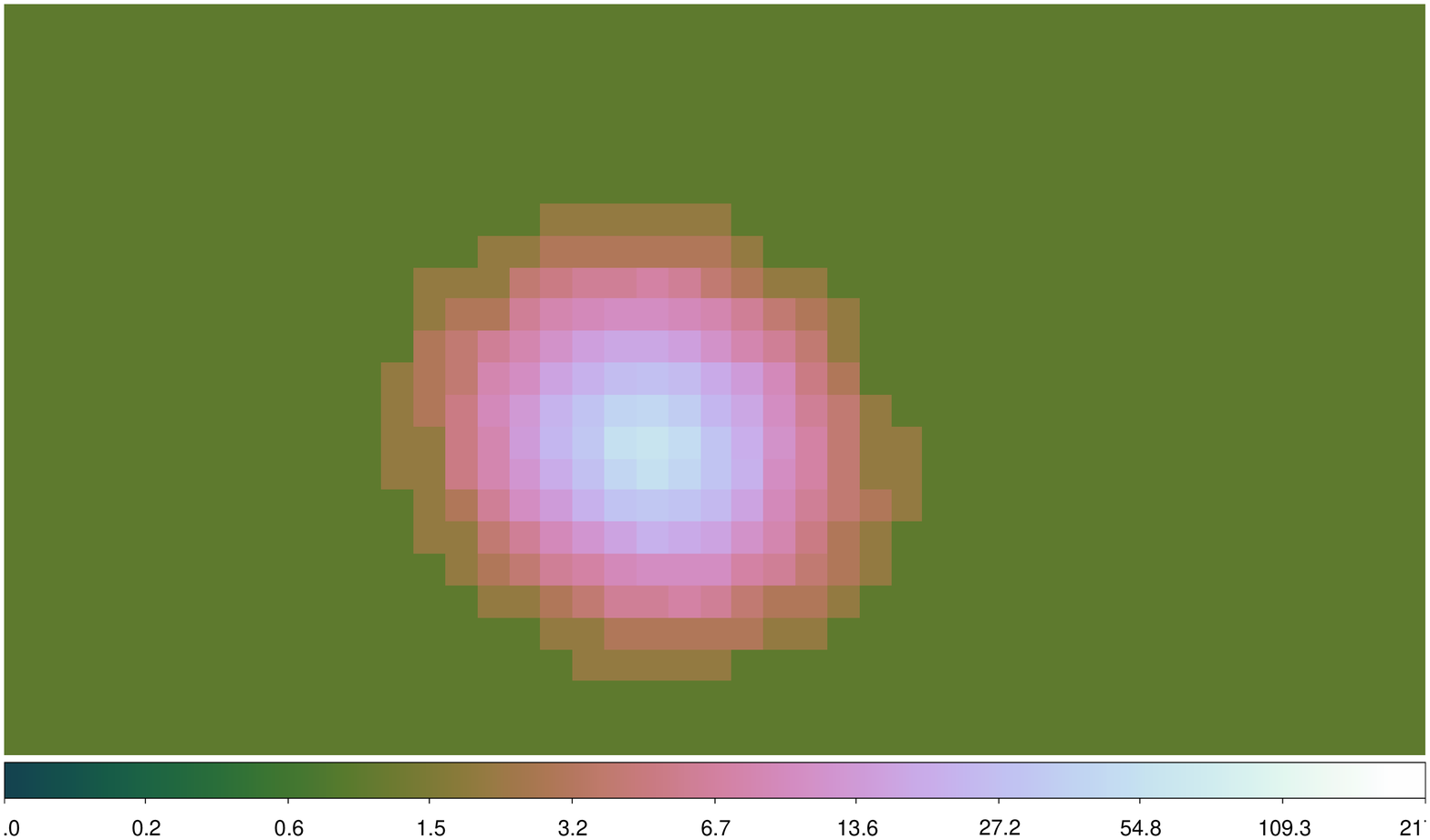}}
\mbox{\includegraphics[width=40mm]{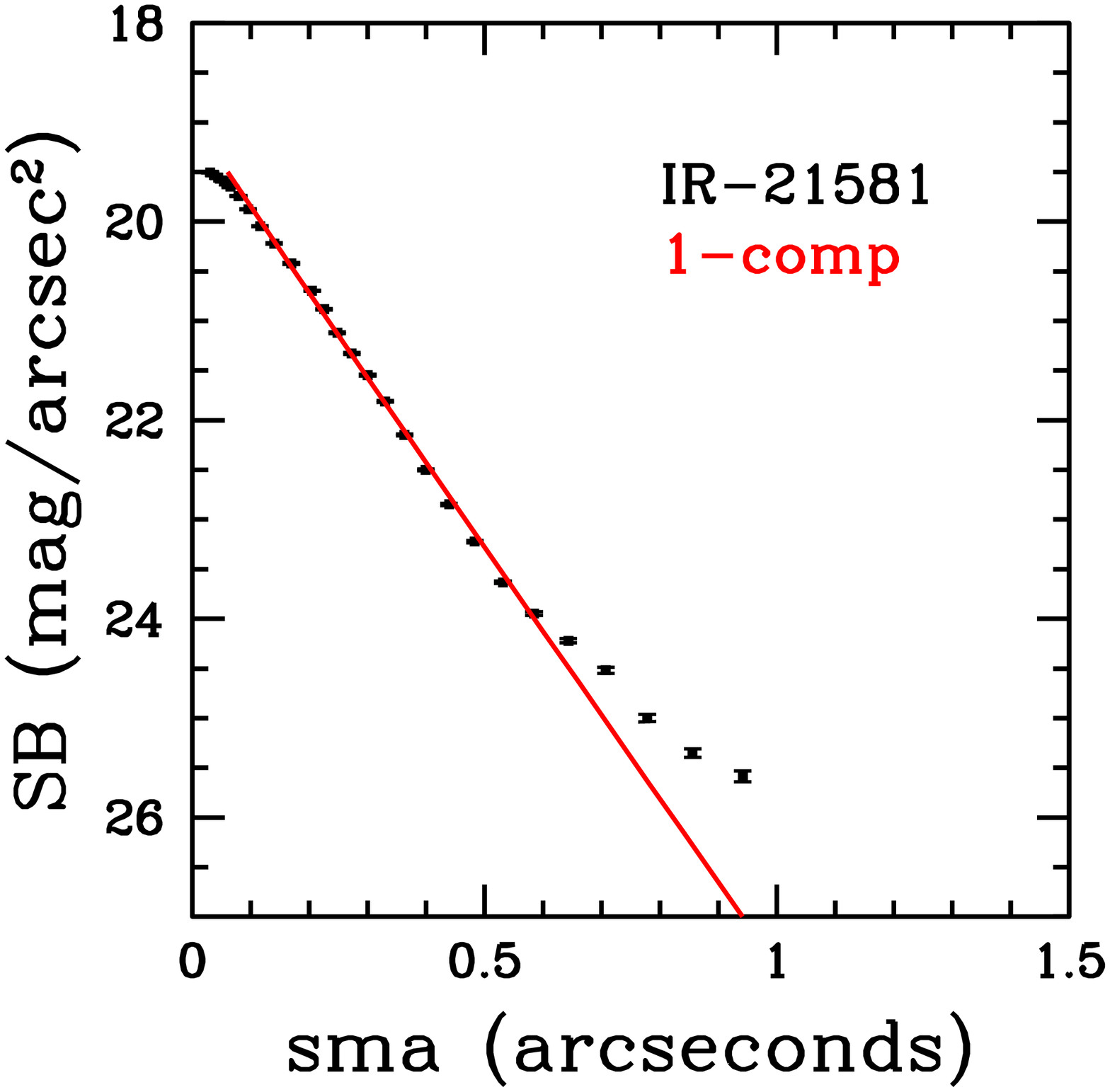}}
\mbox{\includegraphics[width=40mm]{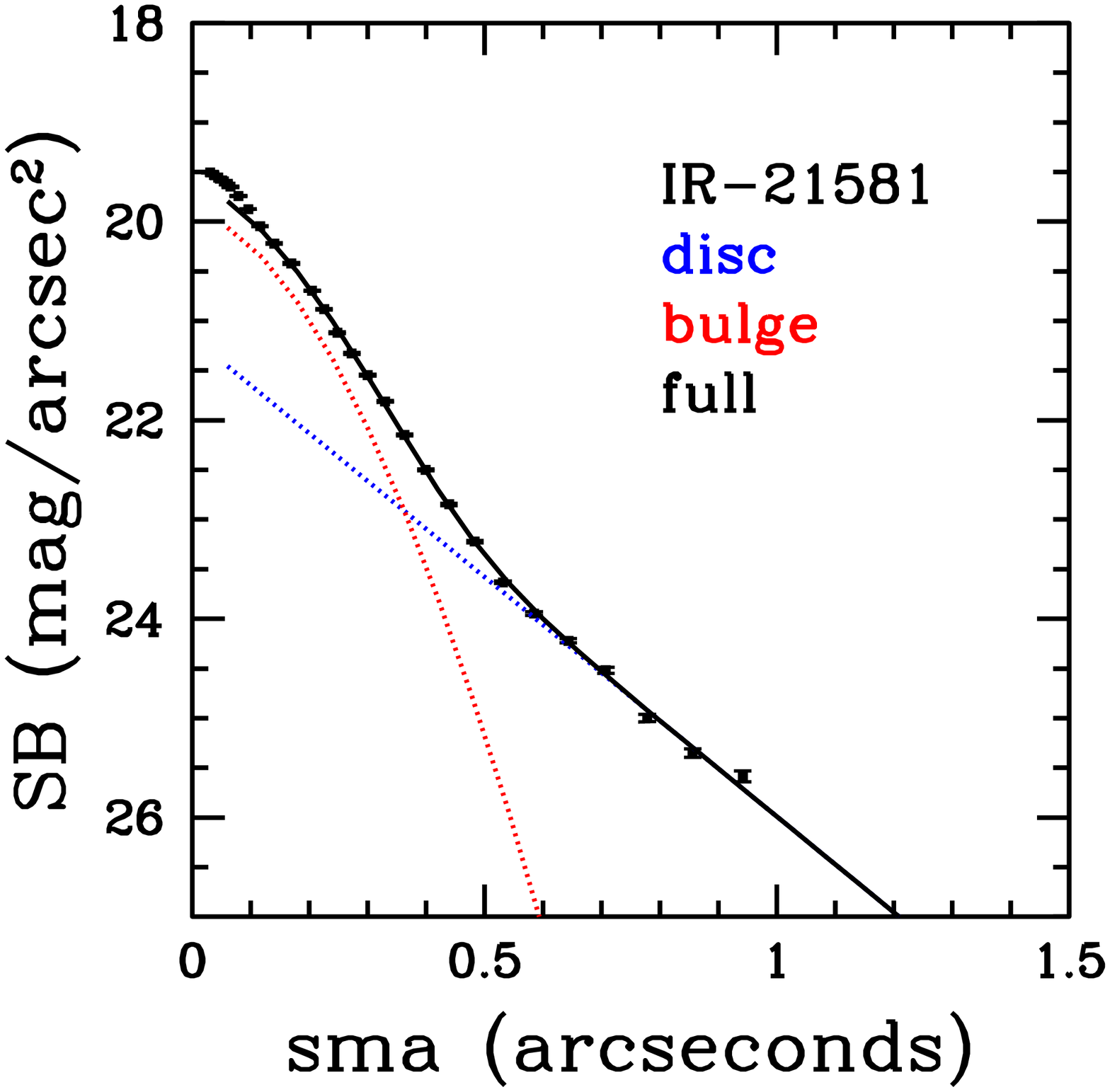}}\\
\mbox{\includegraphics[width=50mm]{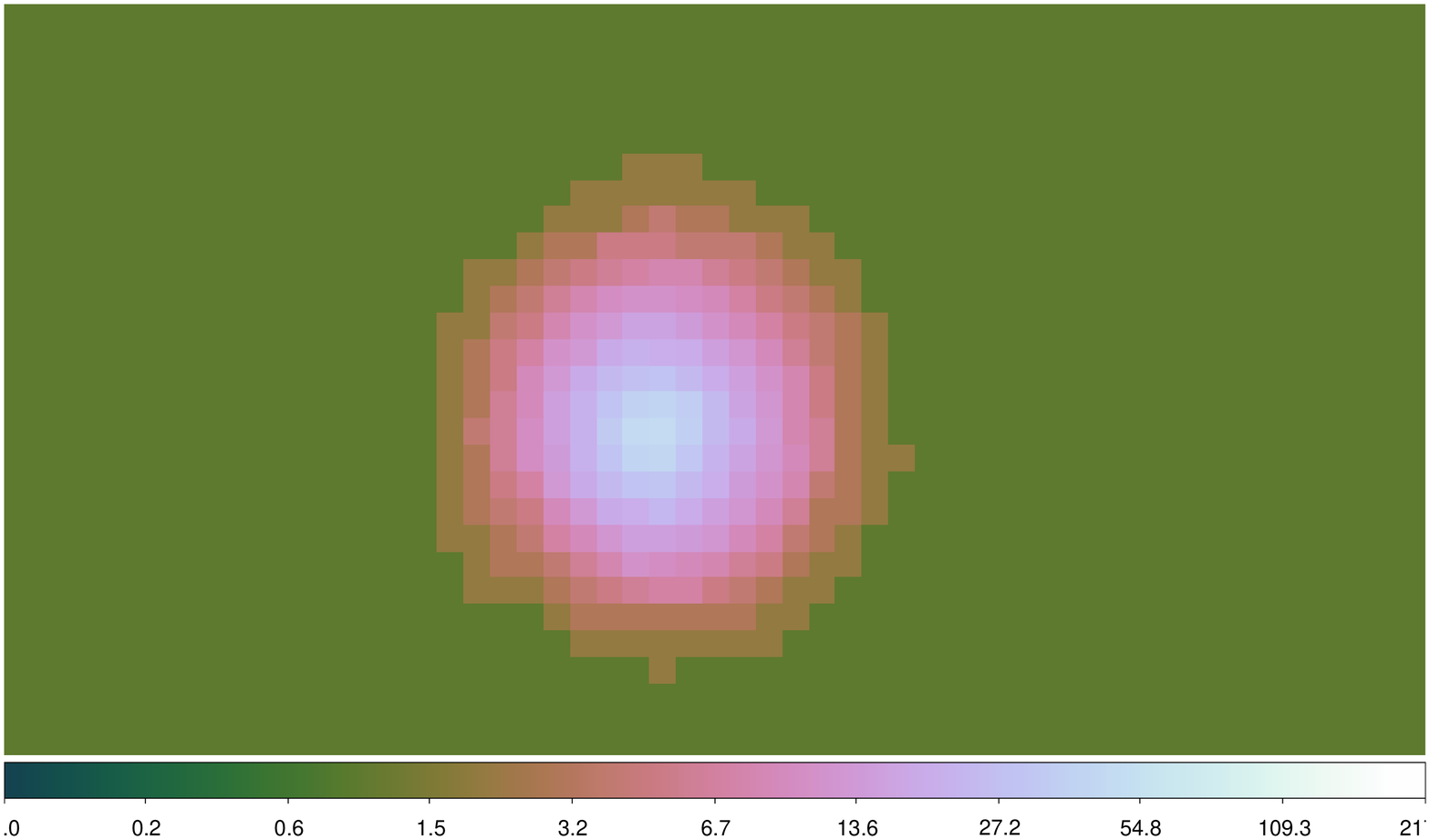}}
\mbox{\includegraphics[width=40mm]{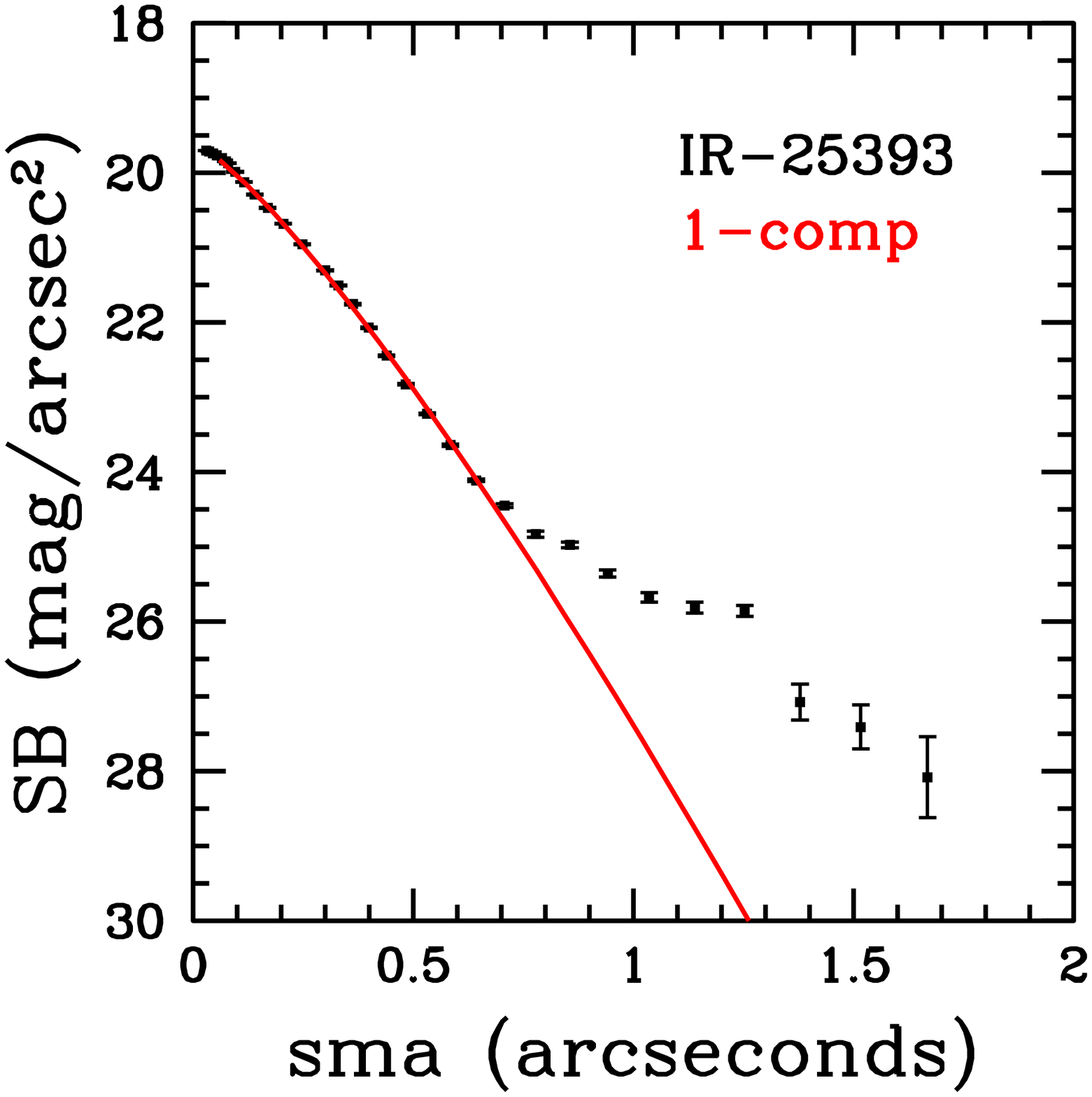}}
\mbox{\includegraphics[width=40mm]{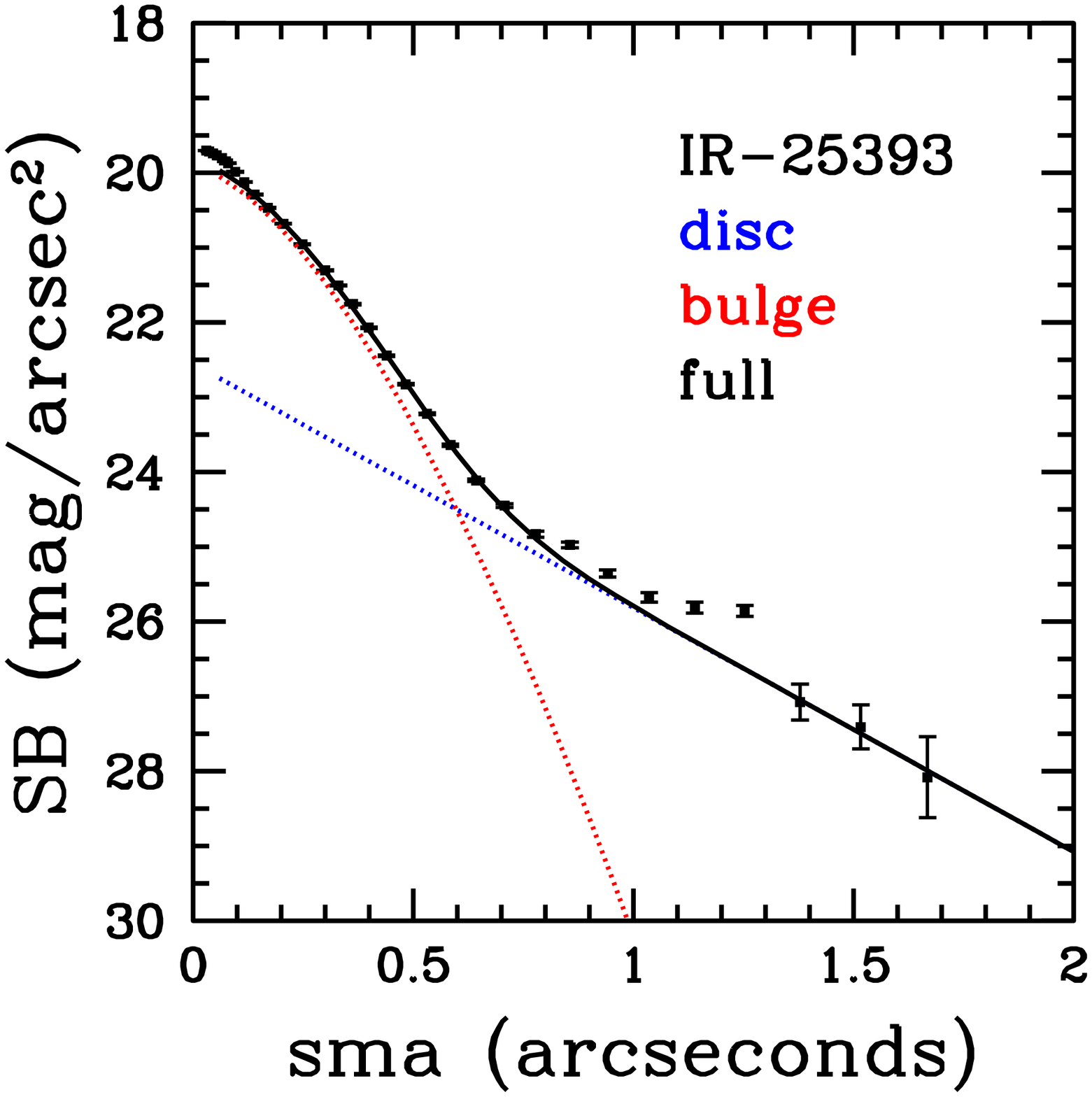}}
\caption{The images and radial profiles of surface brightness are shown for selected disc galaxies with BCBs in rest-frame {\it I}-band. The extent of the profiles is up to that radius till which isophotes converged. In the second column the radial intensity profiles have been fitted with a single S\'ersic function. In the third column they have been fitted with two functions, i.e., the exponential function on the disc part and S\'ersic function on the bulge part. The black solid line represents the sum of the two functions. The redshift and defining parameters of these galaxies, obtained from fitting two functions, are presented in Table~\ref{table-five-selected}.}
\label{selected-profiles}
\end{figure*}

%%%%%%%%%%%%%%%%%%%%%%%%%%%%%%%%%%%%%%%%%%%%%%%%%%%%%%%%%%%%%%%%%%%%%%%%%%

\begin{table*}
\begin{minipage}{180mm}
\caption{Parameters for the five selected disc galaxies with BCBs.}
\begin{tabular}{@{}lllllllllllll@{}}
\hline
ID & $z$ & $R_{eb,B}$ & $<$SB$_{eb,B}$$>$ & $(B/T)_B$ & $R_{eb,I}$ & $<$SB$_{eb,I}$$>$ & $(B/T)_I$ & Mass$_{\ast}$ & SFR & sSFR & $B-I$ & $B-I$\\
 & & kpc & mag/arcsec$^2$ & & kpc & mag/arcsec$^2$ & & (e$+$10)$M_{\odot}$ & $M_{\odot}/yr$ & $1/Gyr$ & disc & bulge\\
\hline
10019 & 0.64 & 2.0 & 17.7 & 0.65 & 2.4 & 16.9 & 0.81 & 7.41 & 3.47 & 0.05 & 0.3 & 1.2\\
17096 & 0.67 & 1.9 & 18.1 & 0.52 & 1.9 & 17.7 & 0.67 & 1.05 & 40.98 & 3.91 & 0.3 & 1.1\\
19985 & 0.97 & 2.4 & 18.4 & 0.38 & 2.1 & 17.6 & 0.64 & 4.47 & 29.99 & 0.67 & 0.3 & 1.2\\
21581 & 0.81 & 1.2 & 18.1 & 0.61 & 1.4 & 18.1 & 0.58 & 1.20 & 3.64 & 0.30 & 0.9 & 0.4\\
25393 & 0.94 & 1.4 & 17.9 & 0.60 & 1.8 & 17.9 & 0.77 & 0.85 & 26.33 & 3.09 & 0.0 & 0.5\\
\hline
\label{table-five-selected}
\end{tabular}
\end{minipage}
\end{table*}

%%%%%%%%%%%%%%%%%%%%%%%%%%%%%%%%%%%%%%%%%%%%%%%%%%%%%%%%%%%%%%%%%%%%%%%%%%%%%% 

\begin{table*}
\begin{minipage}{180mm}
\caption{Median and Median Absolute Deviation for discs with the three bulge types.}
\begin{tabular}{@{}llllllllll@{}}
\hline
Bulge & $R_{eb,B}$ & $<SB_{eb,B}>$ & $(B/T)_B$ & $R_{eb,I}$ & $<SB_{eb,I}>$ & $(B/T)_I$ & Mass$_{\ast}$ & SFR & sSFR\\
type & kpc & mag/arcsec$^2$ & & kpc & mag/arcsec$^2$ & & (e$+$10)$M_{\odot}$ & $M_{\odot}/yr$ & $1/Gyr$\\
\hline
pseudo & 2.9($\pm$0.9) & 21.3($\pm$0.4) & 0.12($\pm$0.05) & 3.3($\pm$0.8) & 20.4($\pm$0.6) & 0.44($\pm$0.17) & 0.48($\pm$0.33) & 4.75($\pm$2.67) & 0.97($\pm$0.40)\\
classical & 2.6($\pm$0.7) & 20.1($\pm$0.5) & 0.25($\pm$0.08) & 2.7($\pm$0.7) & 19.5($\pm$0.4) & 0.62($\pm$0.17) & 0.60($\pm$0.34) & 6.51($\pm$3.54) & 1.27($\pm$0.55)\\
BCB & 2.0($\pm$0.5) & 18.8($\pm$0.3) & 0.42($\pm$0.10) & 2.3($\pm$0.4) & 18.6($\pm$0.6) & 0.62($\pm$0.16) & 0.87($\pm$0.58) & 14.12($\pm$10.37) & 1.77(0.78)\\
\hline
\label{table-medians}
\end{tabular}
\end{minipage}
\end{table*}

%%%%%%%%%%%%%%%%%%%%%%%%%%%%%%%%%%%%%%%%%%%%%%%%%%%%%%%%%%%%%%%%%%%%%%%%%%%%%%%%%%%%%

\subsection{Robustness of measurements}

Since the selection of BCBs is based on their being above the 3-sigma boundary for ellipticals, it is imperative to ensure that the measurements are robust, i.e., their presence is not due to dispersion in the measurement of classical bulge parameters. Towards that, we created models of disc galaxies with a range of bulge to total light ratios (0.1-0.8) and bulge magnitudes (18-24 mag) using the model-mode of Galfit \citep{Pengetal2002}. Following that, we obtained their intensity profiles using IRAF {\it ellipse} isophotes fitting method and fitted them using Gnuplot's Levenberg-Marquardt algorithm, according to our procedure. We found our procedure to be successful in recovering back the input parameters to a second decimal of accuracy in both magnitudes and bulge-to-light ratios. The accuracy is attributed to the fact that we fit each galaxy individually following a procedure which is guided by visual inspection. Since the procedure intuitively divide the fitting into multiple steps, the number of free variables in each step reduces, contributing to the reduction in degeneracy of solutions.

The second decimal of accuracy is for mock galaxies created with smaller ellipticity values, i.e. less than 0.4. For successively increasingly values of ellipticity, we typically recovered parameters with first decimal of accuracy, still well within the stipulated error range. Note that magnitudes and sizes can also affect the accuracy if one component is significantly fainter or shorter than the other. However, our full sample consists of bright galaxies with substantial (well-resolved) bulge ($B/T$$>$0.05) and disc ($B/T$$<$0.90) components. It is also important to acknowledge that mock galaxies are free from noise/contamination (i.e., overlapping clumps, diffuse light from neighbouring sources, etc.) and have well defined symmetrical structure. Given reasonable conditions, our procedure recovers galaxy parameters in a highly consistent manner. Since all galaxies in our sample are imaged and measured under similar conditions, the dispersion of bulges on the Kormendy plane is not due to measurement errors.

%%%%%%%%%%%%%%%%%%%%%%%%%%%%%%%%%%%%%%%%%%%%%%%%%%%%%%%%%%%%%%%%%%%%%%%%%%%%%%%%%%%%%

\subsection{Stellar parameters}

We have obtained the stellar parameters, i.e., stellar mass and star formation rate (SFR), for our sample from 3DHST spectroscopic catalog \citep{Brammeretal2012,Momchevaetal2016}. Here, stellar masses have been obtained using FAST algorithm \citep{Krieketal2009} which fits the extensive range (0.3-8.0 $\mu$m) of photometric data from 3DHST \citep{Skeltonetal2014} with stellar population synthesis templates. The input includes spectroscopic redshifts (measured from simultaneously fitted interlaced 2D spectra and SEDs), a grid of \citet{BruzualandCharlot2003} models that assume a \citet{Chabrier2003} IMF with solar metallicity, a range of ages, exponentially declining SF histories and dust extinction.

Star formation rates have been estimated by \citet{Whitakeretal2014} taking into account the rest-frame UV emission and re-radiated light at Far-IR wavelengths. For rest-frame UV emission, luminosity at 0.28 $\mu$m (rest-frame) is obtained from the best fit template and for Far-IR, flux density at 24 $\mu$m is obtained from Spitzer/MIPS. \citet{Whitakeretal2014} elaborate with empirical evidence that their method of adding up obscured and unobscured stellar light, provides most reliable SFRs.

%%%%%%%%%%%%%%%%%%%%%%%%%%%%%%%%%%%%%%%%%%%%%%%%%%%%%%%%%%%%%%%%%%%%%%%%%%%%%%%%%%%%%%%%%%%%%%%%% 

\subsection{Contamination by AGNs}

It is important to check if the properties of disc galaxies in our sample have got affected by the presence of AGNs inside these galaxies. Towards that, we cross-match our sample with the latest CDFS point-source catalog obtained using 7 Ms Chandra exposure \citep{Luoetal2017}. We match RA Dec of our sample with their catalog allowing for a maximum normal distance of 4 arcseconds or 0.001 degrees. Note that it is a very liberal condition since all our galaxies are significantly smaller than 4 arcseconds in angular size. Other than that, we allow a redshift difference of 0.05 between our sample and their catalog, which is again liberal with regard to redshift accuracies. Even after carrying out such broad probing, we find that only $\sim$2\% of our galaxies (8 out of 358) match with the AGN population. Amongst the 8 identified, 6 belong to the sample of discs with classical bulges and one each belongs to discs with BCBs and pseudo bulges. Note that as the constraints are tightened, none of the galaxies shows an exact match. For these 8, we find that their inclusion does not affect any of the measurements or results in any manner.

%%%%%%%%%%%%%%%%%%%%%%%%%%%%%%%%%%%%%%%%%%%%%%%%%%%%%%%%%%%%%%%%%%%%%%%%%%%%%%%%%%%%%%%%%%%%%%%   
 
\subsection{Local sample for comparison}

A crucial aspect of understanding disc galaxies with BCBs, discovered at intermediate redshifts ($0.4<z<1.0$), is to probe the presence of their local counterparts, if any. For a thorough probe, it is essential to cover the largest sample available. \citet{Simardetal2011} performed bulge $+$ disc decomposition in {\it g} and {\it r}-band on 11,23,718 galaxies from SDSS DR7. For bulge and disc parameters, we took all the galaxies for which bulge (with a free S\'ersic index) in combination to the exponential disc provided the best fit. Additionally, we applied the condition that $B/T$ (bulge to total light ratio) is more that 0.05, i.e., the bulge is substantial, which is the case with our intermediate redshift sample. For elliptical galaxy parameters, we took all the galaxies for which single S\'ersic function provided the best fit and S\'ersic index was higher than 3.5. Note that the same S\'ersic cut was applied to obtain elliptical galaxies, at intermediate redshift range, in our sample.

Using the parameters in {\it g} and {\it r} band, we obtain rest-frame {\it B}-band parameters for bulges, host discs and elliptical galaxies. We apply the same magnitude cut as that applied on our sample, i.e., $M_B<-20$, where $M_B$ is the total absolute magnitude of the galaxy in rest-frame {\it B}-band. The redshift range is selected as 0.02-0.05. We thus obtain bulge and host disc parameters for 10225 local galaxies and elliptical galaxy parameters for 4728 local galaxies, in rest-frame {\it B}-band. In addition to checking for the presence of local counterparts, we will examine the placement of our sample with respect to these galaxies on the Kormendy plane, size-magnitude plane, etc., to analyze how further our sample is from the known characteristics of local galaxies.

%%%%%%%%%%%%%%%%%%%%%%%%%%%%%%%%%%%%%%%%%%%%%%%%%%%%%%%%%%%%%%%%%

\begin{table}
\centering
\begin{minipage}{100mm}
\caption{F-test statistic and probability values.}
\begin{tabular}{@{}llllllllll@{}}
\hline
ID & red-$\chi^2$ & red-$\chi^2$ & F-Stat & Probability & Counts-diff\\
 & 1-comp & 2-comp & & & in \%\\
\hline
10019 & 10.8 & 6.7 & 4.2 & 4.65e-03 & 0.1\\
17096 & 59.9 & 3.2 & 85.5 & 2.37e-12 & 4.9\\
19985 & 81.4 & 6.8 & 62.2 & 1.35e-12 & 2.1\\
21581 & 88.7 & 8.1 & 64.4 & 3.79e-09 & 11.0\\
25393 & 72.8 & 6.4 & 19.4 & 9.44e-06 & 1.1\\
\hline
\label{table-ftest}
\end{tabular}
\end{minipage}
\end{table}

%%%%%%%%%%%%%%%%%%%%%%%%%%%%%%%%%%%%%%%%%%%%%%%%%%%%%%%%%%%%%%%%%%%%%%%%%%%%%%%%%%

\subsection{Selecting few BCBs}

Out of the 43 disc galaxies with BCBs, we select 5 galaxies whose bulges are most bright and most compact in both rest-frame {\it B} and {\it I}-band. For these 5 galaxies, the average effective surface brightness of the bulge, in both rest bands ($<$$SB_{eb,B}$$>$ and $<$$SB_{eb,I}$$>$), is found to be less than $18.5$ mag/arcsec$^2$ and effective radius of the bulge, in both bands ($R_{eb,B}$ and $R_{eb,I}$), is less than $2.5$ kpc. In Table~\ref{table-five-selected} the parameters of the selected galaxies are given for both rest bands. Note that stellar mass (Mass$_{\ast}$), star formation rate (SFR) and specific star formation rate (sSFR) are for the full galaxy, i.e., BCB $+$ host disc. These sources are also marked separately in Fig.~\ref{kormendy-selection}. Presenting the study on disc galaxies with BCBs, our focus will be more on these sources whose bulges are furthest in properties from the known bulge types.

%%%%%%%%%%%%%%%%%%%%%%%%%%%%%%%%%%%%%%%%%%%%%%%%%%%%%%%%%%%%%%%%%%%%%%%%%%%%%%%%%%%%%%%%%%%%%%%%%%%%%%%%%%%%%%%%%%%%%%%%%%%%%%%%%%%%%%%%%%%%%%%%%%%

\begin{figure*}
\mbox{\includegraphics[width=65mm]{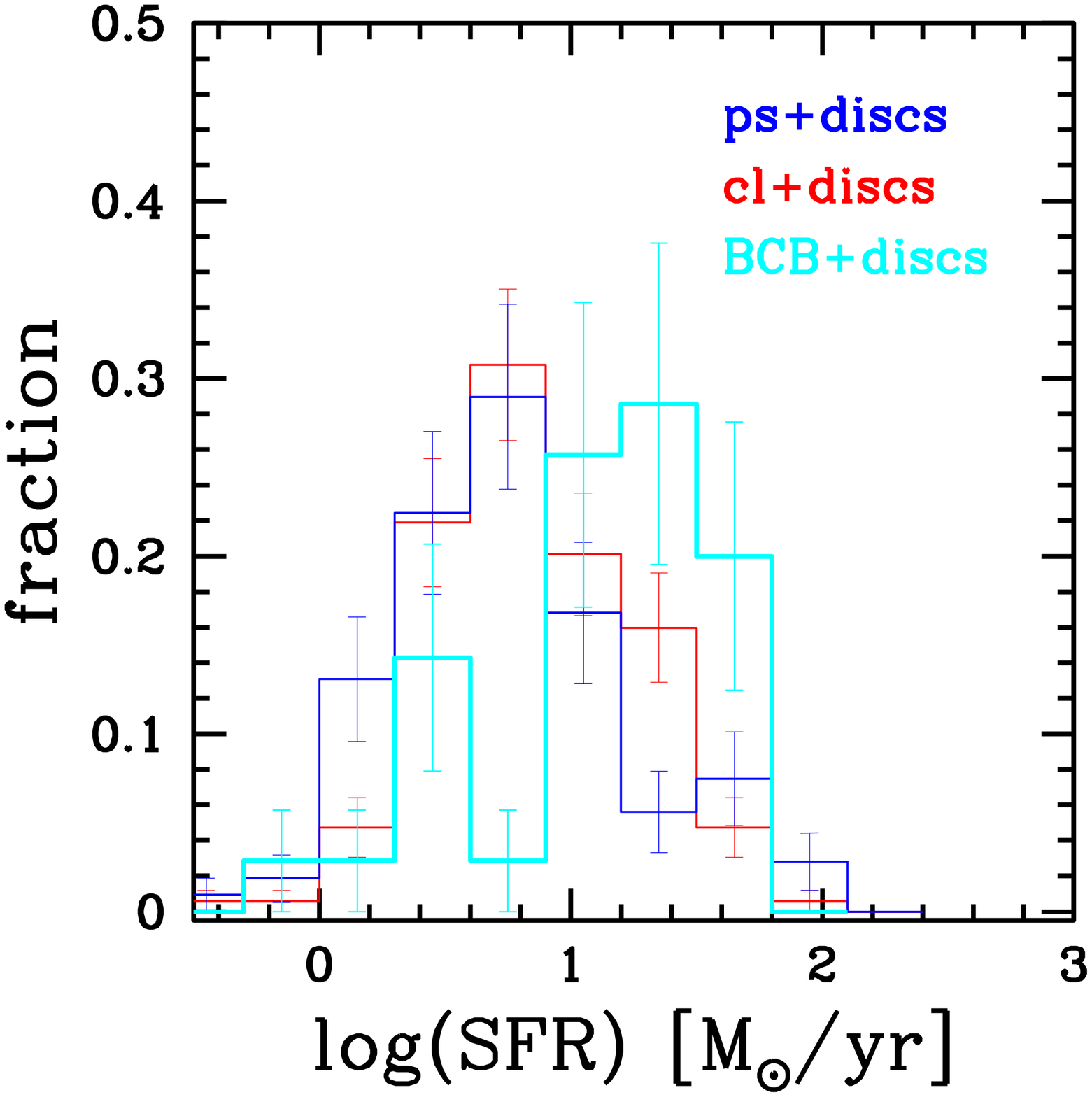}}
\mbox{\includegraphics[width=65mm]{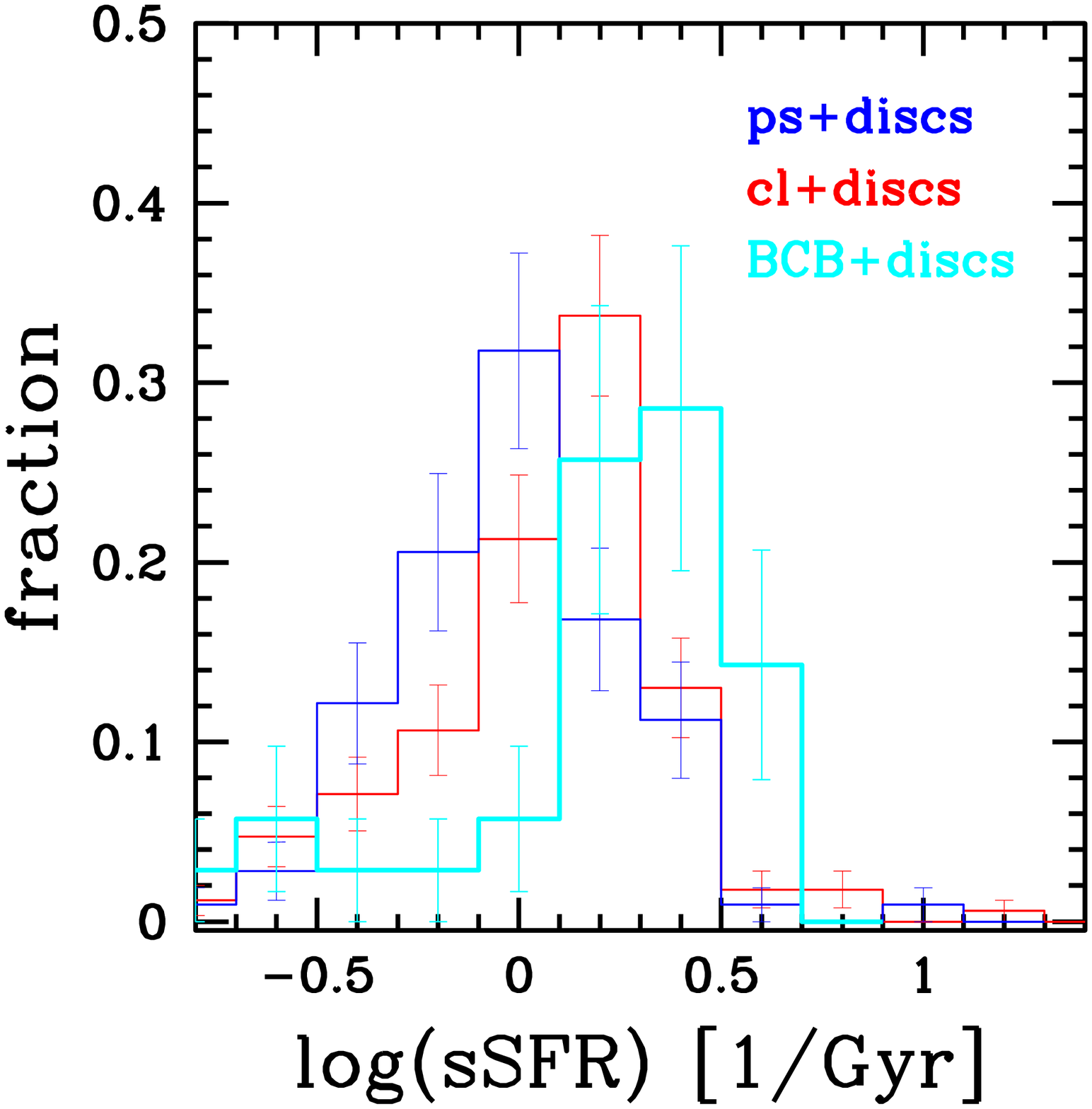}}\\
\mbox{\includegraphics[width=65mm]{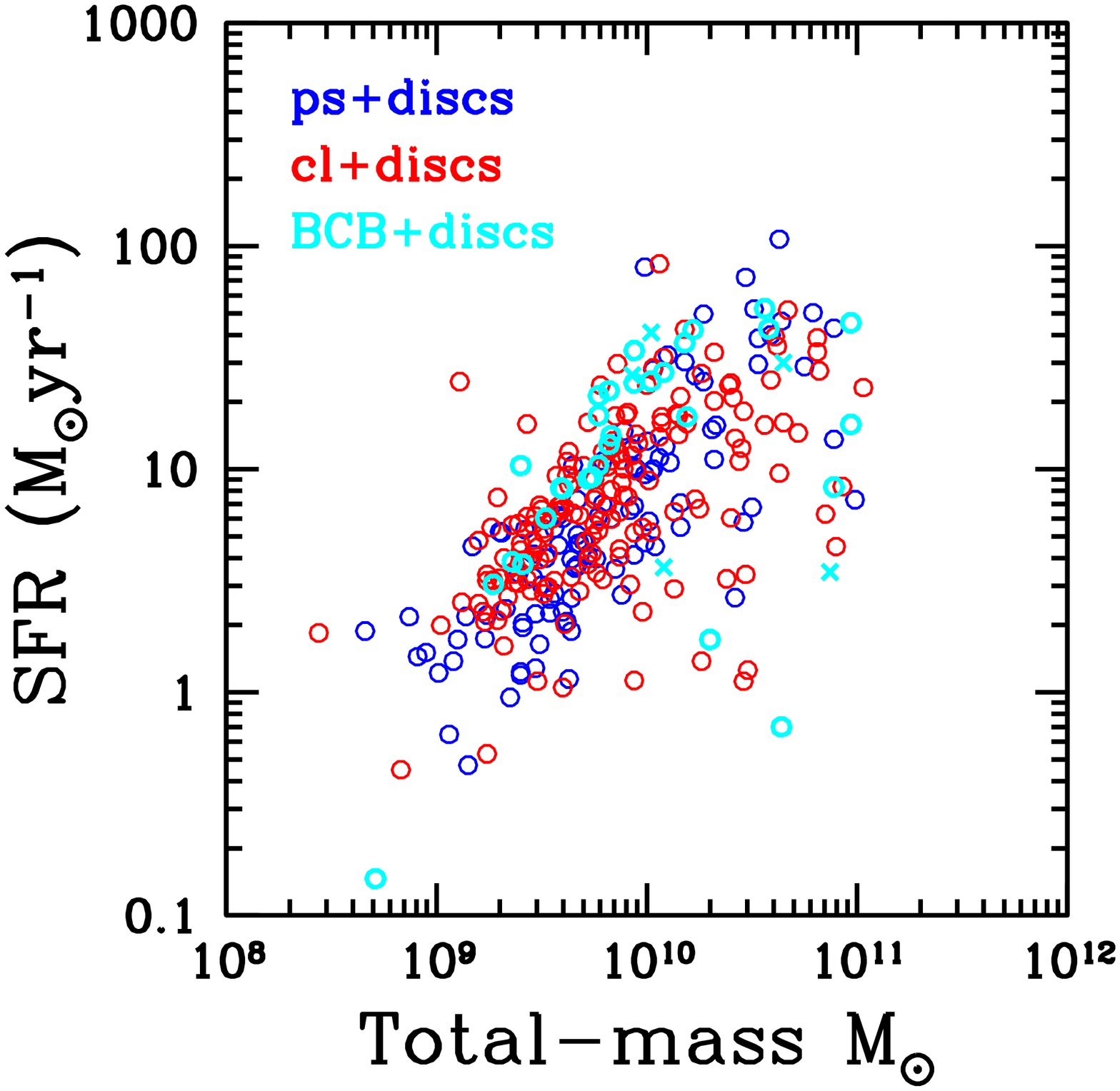}}
\mbox{\includegraphics[width=65mm]{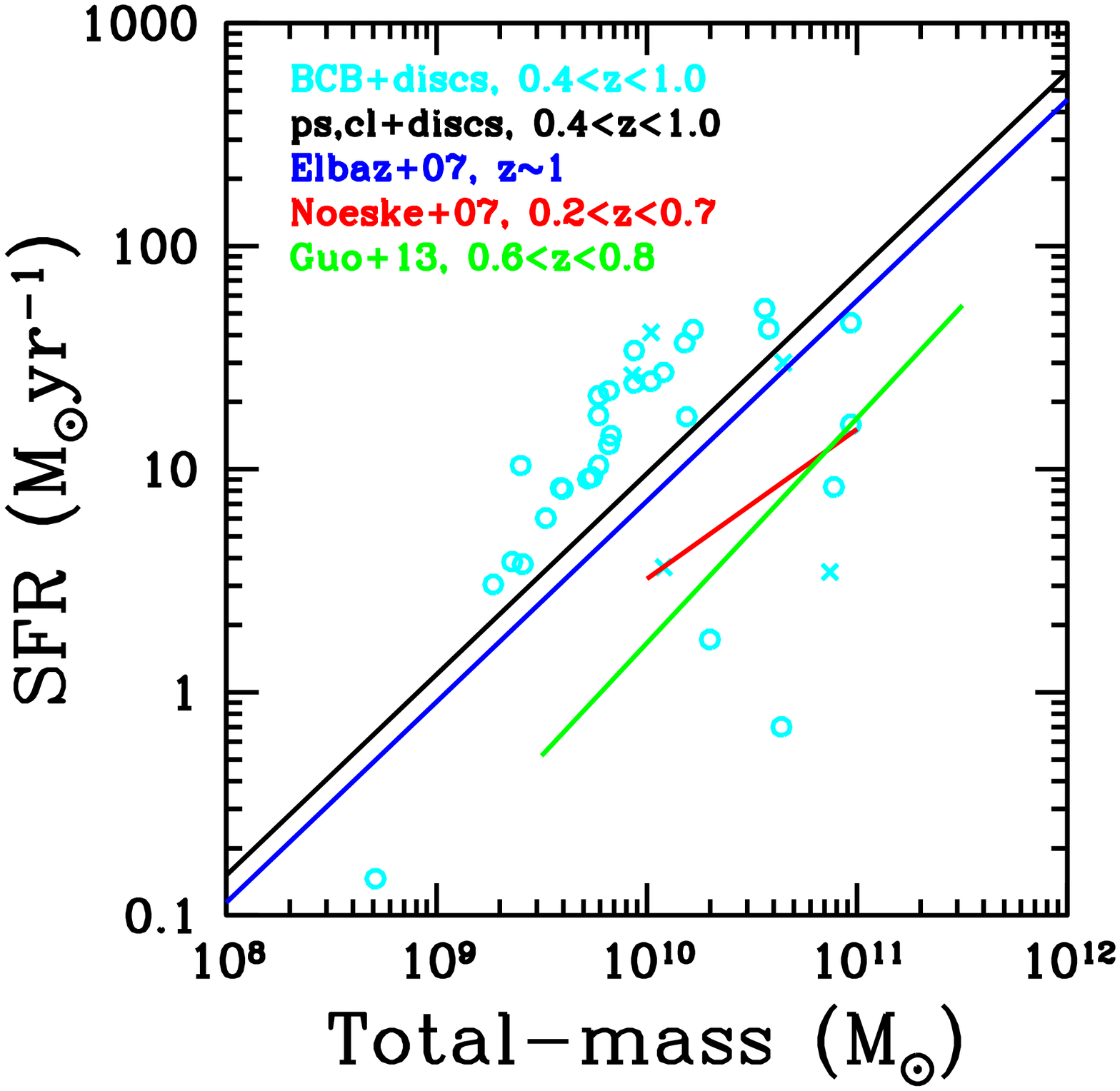}}
\caption{{\it Top panels:} Histogram of the star formation rate (SFR) and specific star formation rate (sSFR) is shown for disc galaxies with pseudo (ps), classical (cl) and BCB bulges. {\it Bottom-Left panel} shows the placement of disc galaxies with the three bulge types on the main-sequence, i.e, SFR-stellar mass plane. {\it Bottom-Right panel} shows the same plot with only disc galaxies with BCBs. For other disc galaxies in our sample, i.e., those with pseudo or classical bulges, relation is marked with a solid black line. Other known relations for star forming galaxies, from the literature, at similar redshifts, are also shown in various colours.}
\label{histsfr-and-relations}
\end{figure*}

%%%%%%%%%%%%%%%%%%%%%%%%%%%%%%%%%%%%%%%%%%%%%%%%%%%%%%%%%%%%%%%%%%%%%%%%%%%%%%%%%%%%%

\section{Results}

\subsection{Distinctness of BCBs}

Table~\ref{table-medians} tabulates the median and corresponding median absolute deviation of the parameters of the galaxies with three bulge types. BCBs are $\sim1$ mag brighter than classical bulges and $\sim2$ mag brighter than pseudo bulges, in both rest-frame {\it B} and {\it I}-band. Their sizes are half a kpc shorter than classical bulges and $\sim1$ kpc shorter than pseudo bulges. Other than being brighter and more compact, they are significantly more dominant inside the galaxy. The fraction of total galaxy light inside a BCB is $\sim2$ times larger than that inside a classical bulge and $\sim4$ times larger than that inside a pseudo bulge. Classical bulges account for $\sim25$\% of galaxy's total light in optical ($(B/T)_B$), which is same as that found by \cite{Tascaetal2014} for their intermediate redshift sample. BCB bulges, on the other hand, account for $\sim42$\% (median value) of galaxy's total light.

This ratio is even larger for our selected few galaxies (see Table~\ref{table-five-selected}), where for rest-frame {\it I}-band, $(B/T)_I$ reaches 70-80\%. For such galaxies, more than others, it is imperative to confirm the presence of a disc, i.e., confirm the requirement of fitting the disc component. In Fig.~\ref{selected-profiles} we present the images and radial surface brightness profiles, fitted with 1-component as well as our 2-component method, of these selected disc galaxies with BCBs in rest-frame {\it I}-band. As can be seen, fitting profiles with only a S\'ersic component invariably leaves about $\sim35-40$\% of the counts (or light) unaccounted for, mostly in the outer parts of the galaxy.

On the other hand, our two component fitting method accounts for about $\sim90-95$\% of the counts across the profile. Table~\ref{table-ftest} includes a column ``Counts-diff in \%" which provides the percentage of counts difference between the 2-component model profile and observed profile, for each selected galaxy. The table also includes F-test results which provide astounding physical significance for going for 2-component fitting over one component for each source. Reduced $\chi^2$ values for one and two component fitting also signifies the presence of a disc component.

%%%%%%%%%%%%%%%%%%%%%%%%%%%%%%%%%%%%%%%%%%%%%%%%%%%%%%%%%%%%%%%%%%%%

\begin{figure*}
\mbox{\includegraphics[width=55mm]{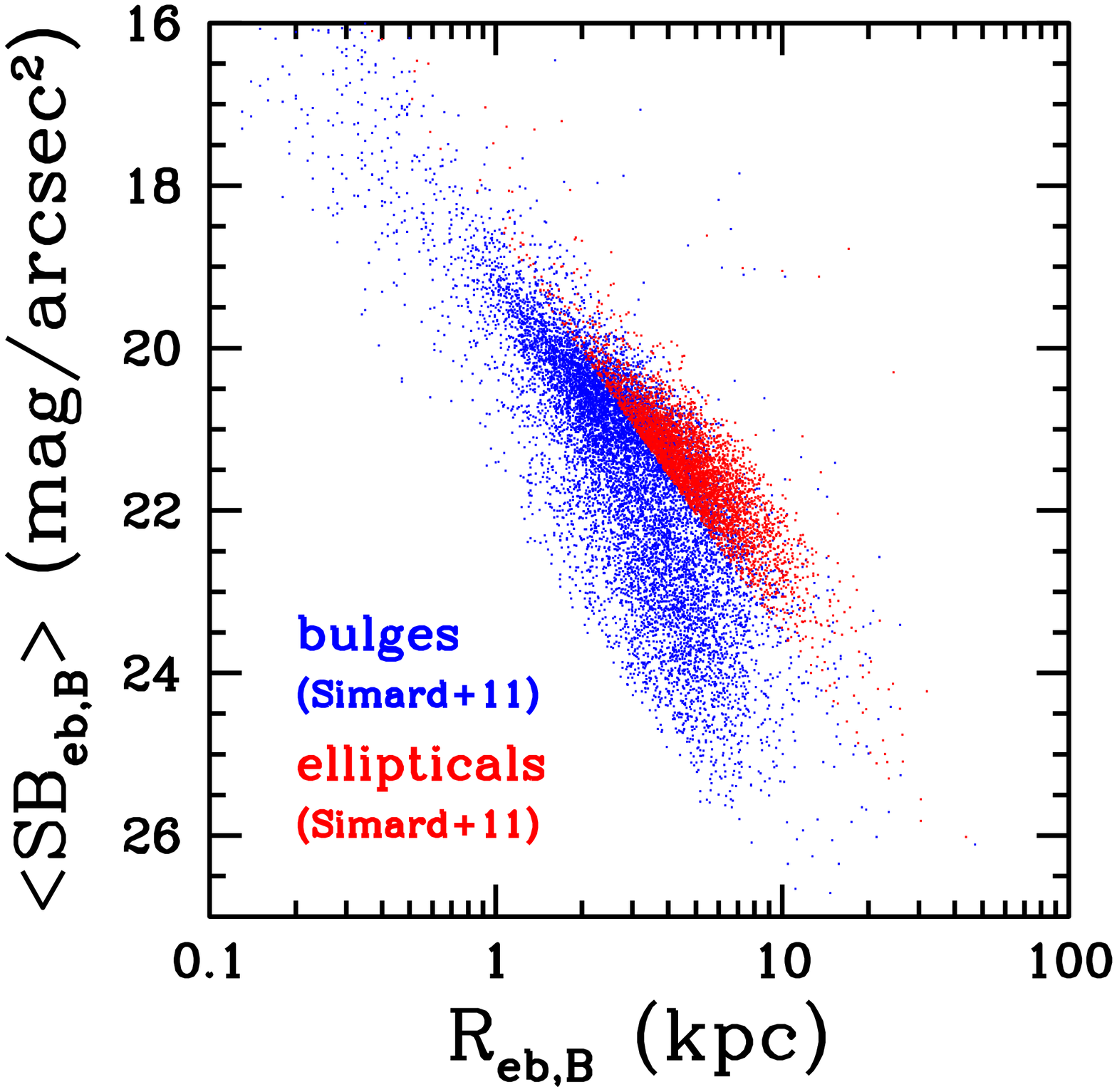}}
\mbox{\includegraphics[width=55mm]{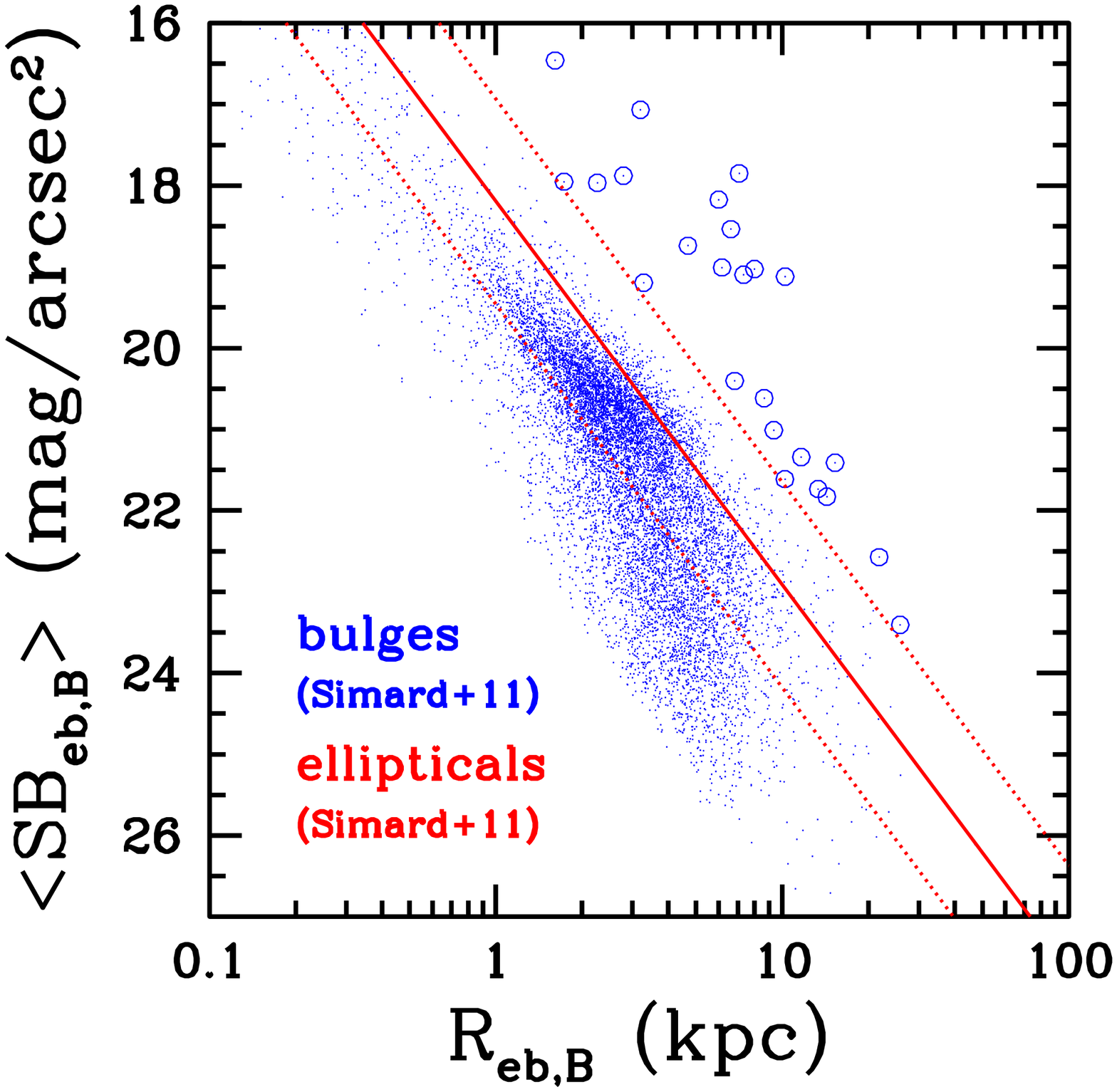}}
\mbox{\includegraphics[width=55mm]{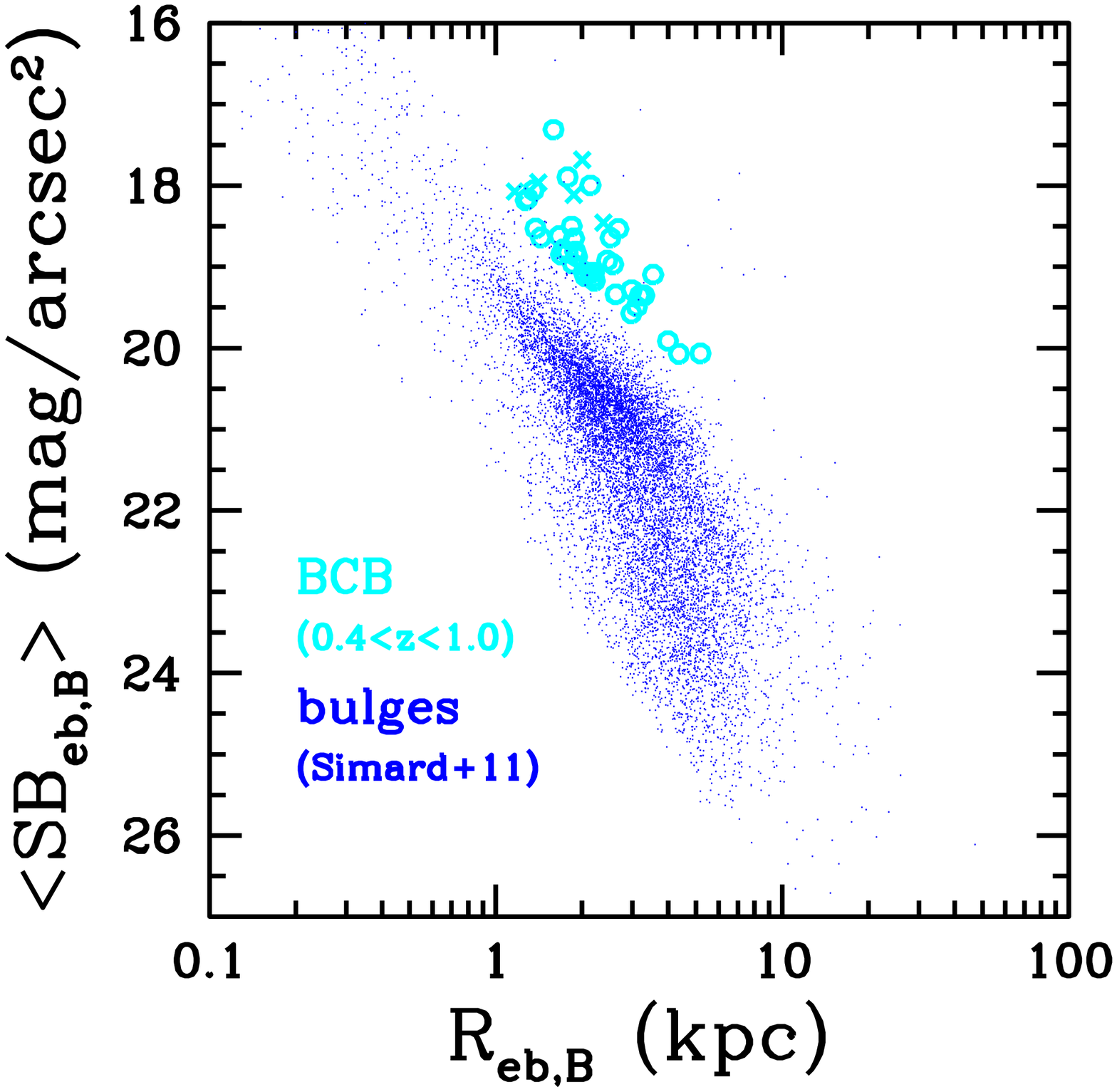}}
\caption{{\it Left panel} shows the placement of local bulges and local elliptical galaxies, both obtained from \citet{Simardetal2011}, on the Kormendy plane. {\it Middle panel} shows the same plot. The local ellipticals are now replaced with their relation marked by the solid red line. The two dashed red lines mark the 3-sigma boundaries of that relation. All local bulges are within or below the 3-sigma boundaries. Only 0.2\% of the local bulges are spotted above the 3-sigma boundary, they have been marked with open-circles. {\it Right panel:} The distribution of our intermediate redshift sample of BCB bulges is shown with respect to all the local bulges on the Kormendy plane.}
\label{comparison-simard}
\end{figure*}

%%%%%%%%%%%%%%%%%%%%%%%%%%%%%%%%%%%%%%%%%%%%%%%%%%%%%%%%%%%%%%%%%%%%%%%%%%%%%%%%%%

In addition to BCBs being more dominant than other bulges, disc galaxies with BCBs are about twice as massive as disc galaxies with pseudo and classical bulges. Consequently, BCBs are $\sim2.5$ times more massive than classical bulges and $\sim6$ times more massive than pseudo bulges. In a nutshell, BCBs are brighter, more compact, more dominant and more massive than other known bulge types in that redshift range.

Another important distinctness of BCB host galaxies is that their star formation rate (SFR) and specific star formation rate (sSFR) is more than a factor of $\sim1.5-2$ times larger than that of disc galaxies with pseudo and classical bulges (Table~\ref{table-medians}). The upper panels of Fig.~\ref{histsfr-and-relations} depict that while discs with pseudo and classical bulges have overlapping distribution of SFR, disc galaxies with BCBs have a disparate distribution shifted towards higher values. In case of sSFR also, discs with BCBs peak towards higher values. We also examined their position on the main sequence, i.e., SFR-Stellar mass plane, in the lower panels of Fig.~\ref{histsfr-and-relations}, with respect to known relations for star forming galaxies at similar redshifts. We find that while disc galaxies with pseudo and classical bulges fall on the main sequence, most BCB host galaxies are outliers to it, i.e., are more star forming for a given stellar mass.

This high star formation activity in BCB host galaxies appears to be more towards the disc part than the bulge. We computed the $B-I$ colour for the bulge and the host disc separately and found the host disc to be bluer than the bulge. Although it is not considered a sensitive indicator at higher redshifts, the colour index showed a marked difference for bulge and disc values. These values for selected galaxies (i.e., disc galaxies with most bright and compact BCBs) are specified in Table~\ref{table-five-selected}.

%%%%%%%%%%%%%%%%%%%%%%%%%%%%%%%%%%%%%%%%%%%%%%%%%%%%%%%%%%%%%%%%%%%%%%%%%%%%%%%%%%%%%%%

\begin{figure*}
\mbox{\includegraphics[width=65mm]{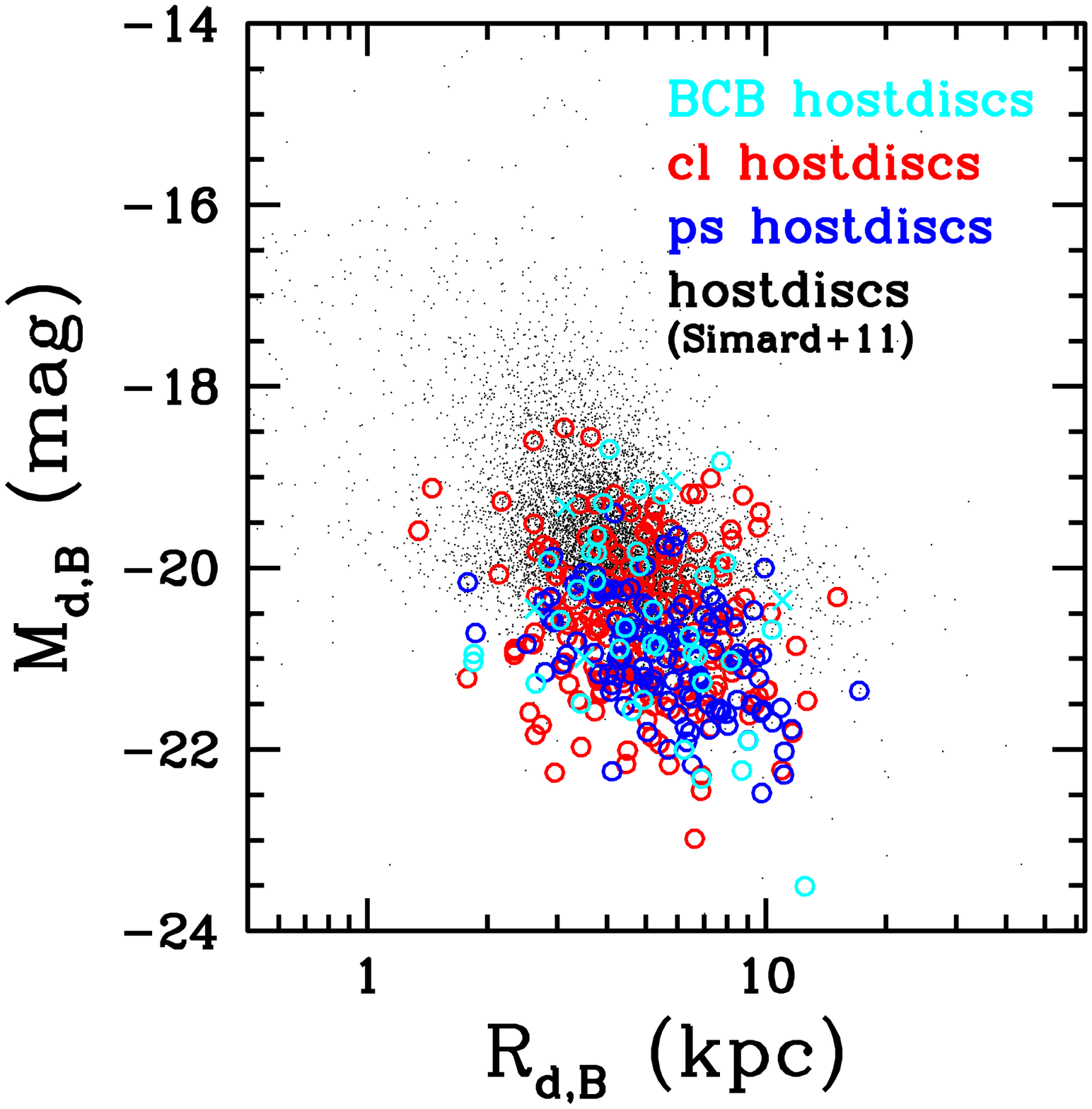}}
\mbox{\includegraphics[width=65mm]{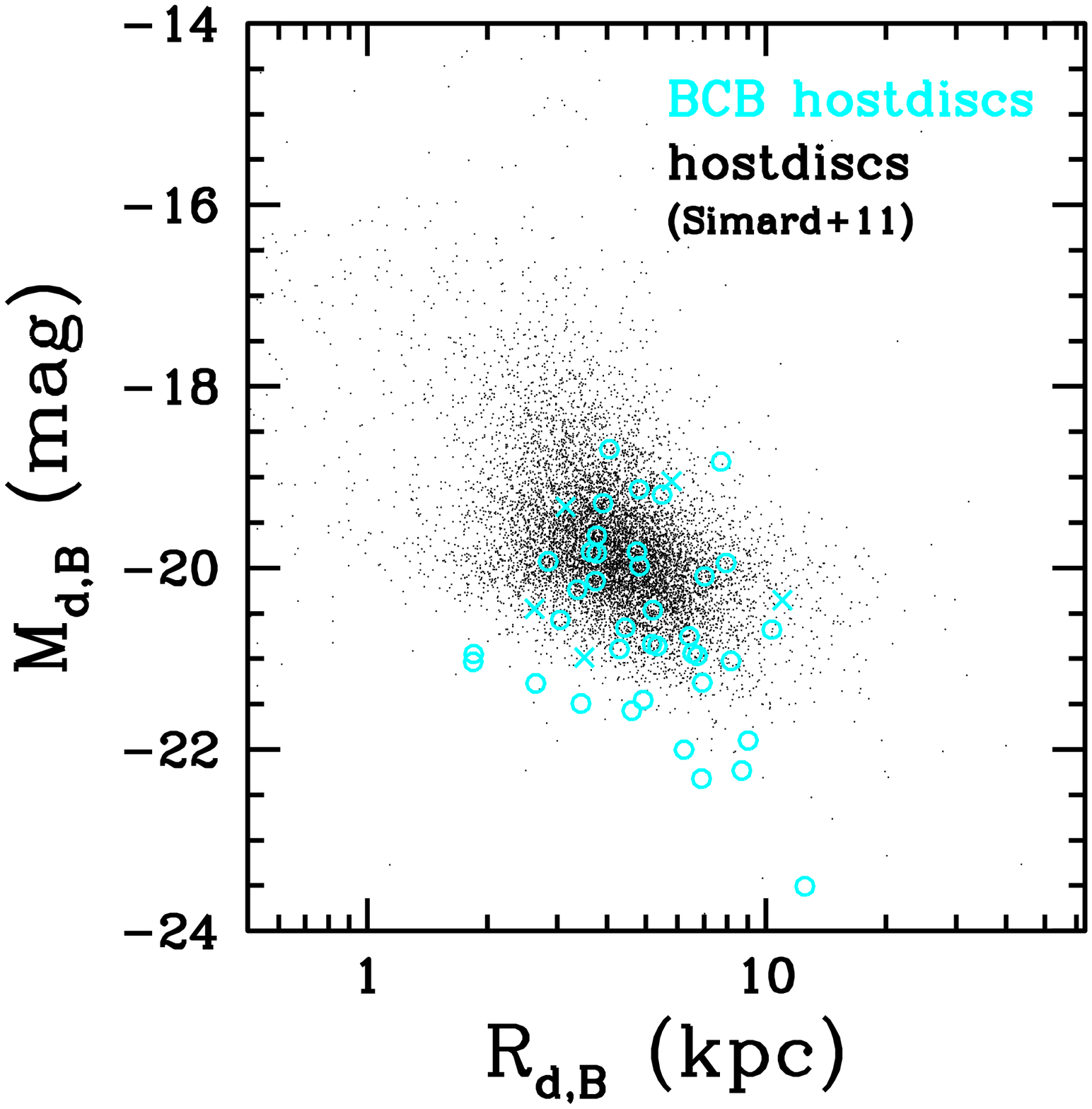}}
\caption{{\it Left panel:} Host discs of pseudo (ps), classical (cl) and BCB are plotted on the size-magnitude plane along with all local host disc parameters obtained from \citet{Simardetal2011}. {\it Right panel} shows the same plot with only BCB host discs and local host discs. Host discs of selected BCBs are marked as crosses in both the plots.}
\label{comparison-hostdisc}
\end{figure*}

%%%%%%%%%%%%%%%%%%%%%%%%%%%%%%%%%%%%%%%%%%%%%%%%%%%%%%%%%%%%%%%%%%%%%%%%%%

\begin{figure*}
\mbox{\includegraphics[width=55mm]{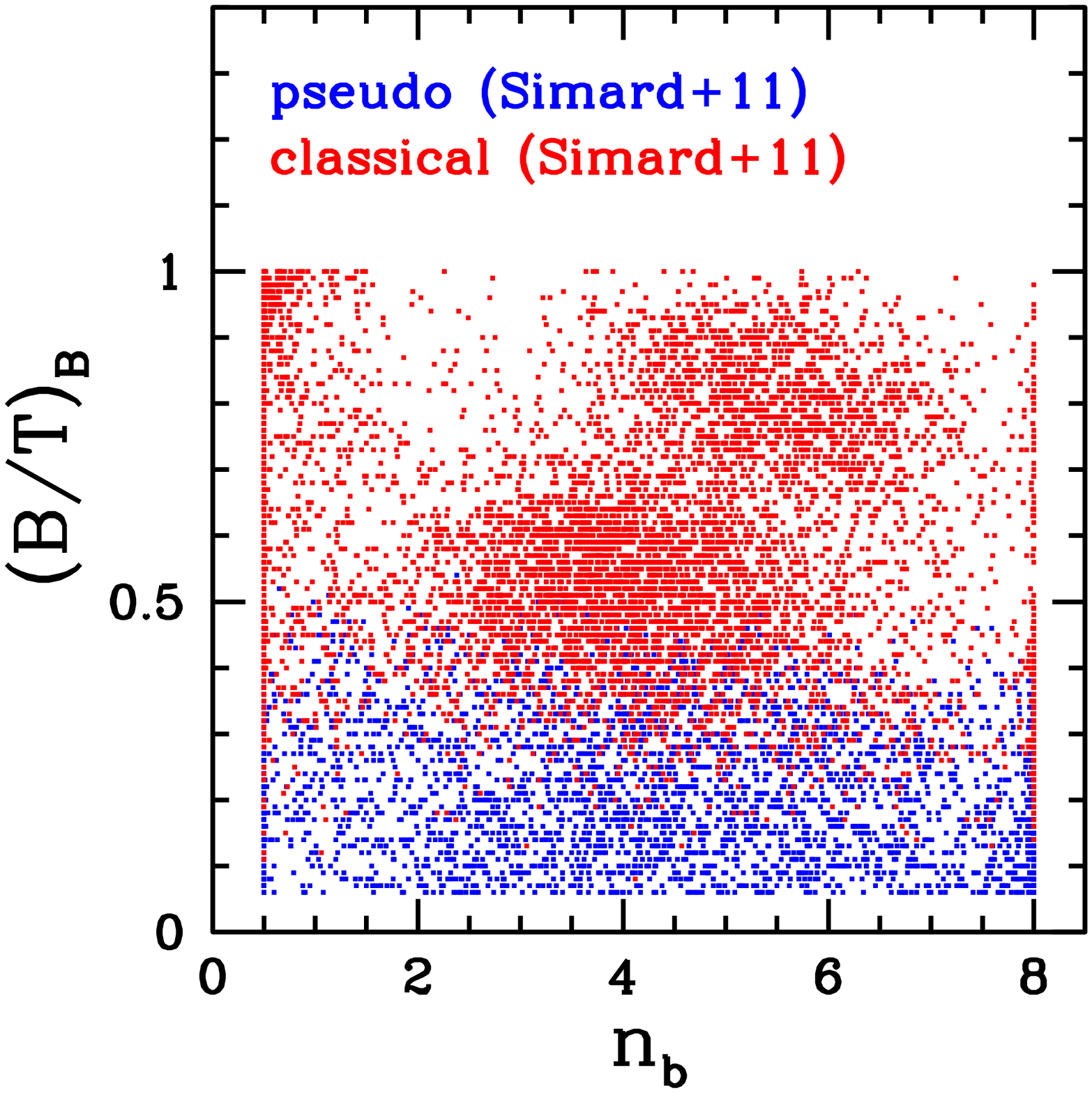}}
\mbox{\includegraphics[width=55mm]{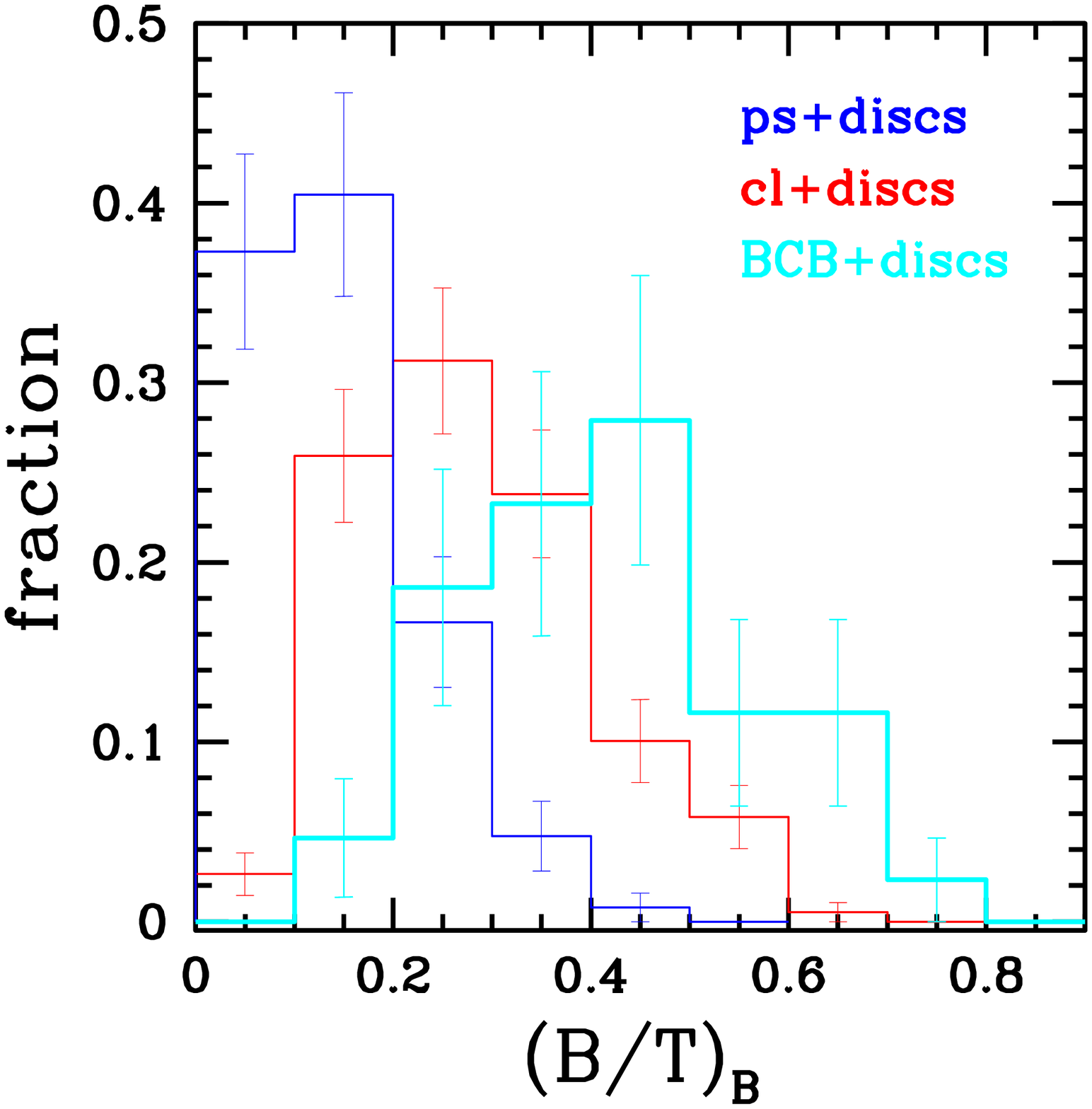}}
\mbox{\includegraphics[width=55mm]{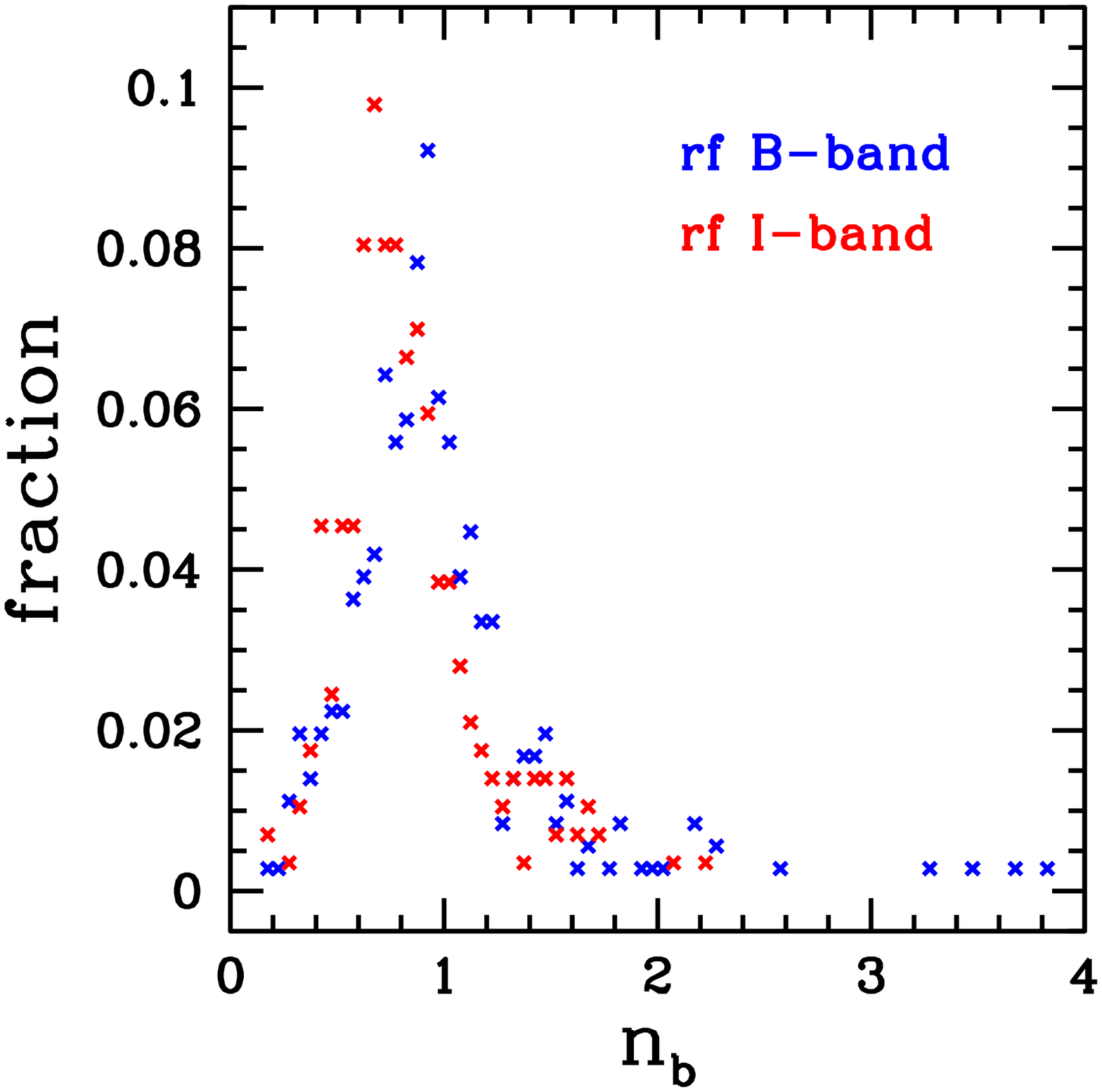}}
\caption{{\it Left panel:} Local bulges, classified according to their placement on the Kormendy-plane, defined by local ellipticals (shown in Fig.~\ref{comparison-simard}), are distributed according to their S\'ersic index and bulge to total-light ratios. {\it Middle panel} shows the distribution of bulge to total-light ratios for our intermediate redshift sample of disc galaxies with pseudo (ps), classical (cl) and BCB bulge types. {\it Right panel} shows the distribution of bulge S\'ersic indices for our full intermediate redshift sample in rest-frame (rf) {\it B} and {\it I}-band.}
\label{sersic-and-btratio}
\end{figure*}

%%%%%%%%%%%%%%%%%%%%%%%%%%%%%%%%%%%%%%%%%%%%%%%%%%%%%%%%%%%%%%%%%%%%%%%%%%%%%%%%%%%%%%%

\subsection{Lack of local counterparts}

The selection of BCBs is according to the Kormendy relation followed by elliptical galaxies at intermediate redshifts, where these bulges, being brighter and compact, emerge as outliers. To examine if there are bulges which can be classified as BCBs in the local Universe, we study the distribution of the entire local bulge and local elliptical galaxy population on the Kormendy plane, Fig.~\ref{comparison-simard}. As shown in the figure, very few local bulges (24 out of 10225, $\sim$0.2\%) reside above the 3-sigma boundary, i.e., are brighter and more compact, than local ellipticals. Thus, there is a near-absence of bulges which can be classified as BCBs in the local Universe.

Fig.\ref{comparison-simard} also shows the distribution of our BCB sample with respect to the entire local bulge sample. Unlike pseudo and classical bulges, BCB bulges show nearly no overlap with the local bulge population, suggesting a morphological transformation of this bulge population from intermediate redshifts to the present epoch.

We also examined if the host discs of BCBs show dissimilar behaviour, in terms of their luminosity or size, with respect to host discs of other bulges and local host discs. In Fig.~\ref{comparison-hostdisc}, host disc parameters of the three bulges are plotted on the size-magnitude plane along with all local host disc parameters obtained from \citet{Simardetal2011}. Interestingly, all host-discs, irrespective of the bulge they host, show overlapping distribution. Also, the distribution does not seem to have evolved with time. Thus, the distinctness of disc galaxies with BCBs and sharp fall in their population with time appears to be solely due to their bulge characteristics.

%%%%%%%%%%%%%%%%%%%%%%%%%%%%%%%%%%%%%%%%%%%%%%%%%%%%%%%%%%%%%%%%%%%%%%%%%%%%%%

\subsection{Other morphology indicators}

We classify the local sample of bulges into classical (those within 3-sigma boundaries, 6465 out of 10225) and pseudo (those below the 3-sigma boundary, 3736 out of 10225) according to the Kormendy relation followed by local ellipticals (Fig.~\ref{comparison-simard}). Examining the distribution of their bulge to total light (B/T) ratios and bulge S\'ersic-indices ($n_b$), in Fig.~\ref{sersic-and-btratio}, we find interesting results. While B/T comes out as a fine demarcator of the two bulge types, there is no distinction in terms of their bulge S\'ersic indices. Although classical bulges show some clustering around $n_b$$=$4 and $n_b$$=$6, pseudo bulge S\'ersic-indices are uniformly scattered. Same is true for our sample as well. While B/T distributions are successively higher for the three bulge types, $n_b$ distributions are overlapping.

Bulge S\'ersic-index is a highly debatable topic in the literature. To obtain insight, \citet{Simardetal2011} selected a subsample of galaxies for which they had good enough images to study bulge profile shapes. Studying the distribution of $n_b$ for this sample, they observed a peak at 0.5-0.55. They found that all the galaxies ($\sim$6700) belonging to this range had effective radius of the bulge between 1-2 arcseconds. They remarked that $n_b$ is measured to be lower for bulges with smaller sizes. This is indeed the case at higher redshifts, where, size of a typical galaxy is about half of its present size. Also, bulge covers a lesser fraction of the total size at higher redshifts than it does in the local. Almost all of our bulges have $n_b$ less than 1.5, Fig.~\ref{sersic-and-btratio}. Note that we did not find any decomposition study at higher (non-local) redshifts which presents the distribution of their bulge S\'ersic indices.

%%%%%%%%%%%%%%%%%%%%%%%%%%%%%%%%%%%%%%%%%%%%%%%%%%%%%%%%%%%%%%%%%%%%%%%%%%%%%%%%%%%%%%%%%%%%%%%%%%%%%%%%%%%%%%%%%%%%%%%%%%%%%%%%%%%%%%%%%%%%%%%%%%%%%%%%%%%%%%%%%%%%%%%%%

\begin{figure}
\mbox{\includegraphics[width=65mm]{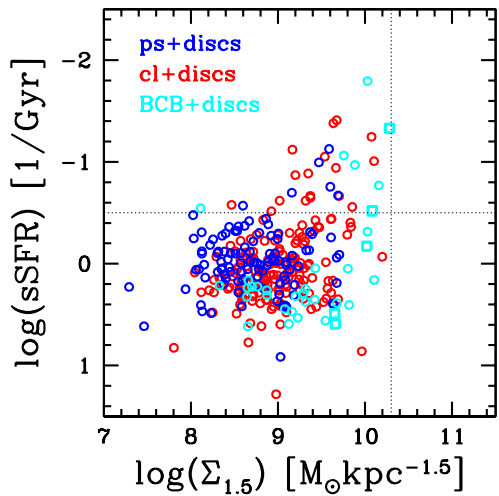}}
\caption{To compare the distribution of our galaxies with respect to those of blue-nuggets, we have plotted them in the plane described in \citet{Barroetal2013}. Here, $\Sigma_{1.5}$ is total stellar mass divided by $R_{e,global}^{1.5}$, where $R_{e,global}$ is effective radius of the full galaxy. The solid lines divide the plot into four quadrants, where, the perpendicular line sets the condition for compactness and horizontal line sets the condition for star-formation. The bottom-right quadrant defines the placement of compact star forming galaxies, i.e., blue nuggets. Although almost all our galaxies are ``star-forming", none are ``compact".}
\label{bluenuggets}
\end{figure}

%%%%%%%%%%%%%%%%%%%%%%%%%%%%%%%%%%%%%%%%%%%%%%%%%%%%%%%%%%%%%%%%%%%%%%%%%%%%%%%%%%%

\section{Conclusions and discussion}
\label{sec:discussion}

Examining the properties of bulges and host discs of bright disc dominated galaxies, at intermediate redshifts, at rest-frame optical and infrared wavelengths, we discover a peculiar class of bulges - referred to as BCBs. These bulges, forming $\sim 12$\% of the total bulge population at these redshifts, are brighter and more compact than elliptical galaxies. BCBs are neither classical nor pseudo-like based on their morphology and star-formation properties. We summarize the major properties of BCBs below:

\begin{itemize}
\item BCBs are $\sim1$ mag brighter than classical bulges and more than $\sim2$ mag brighter than pseudo bulges. 

\item Bulge to total optical light ratio $(B/T)_B$ of BCB host disc galaxies is $\sim0.42$ - this is a factor of $\sim2$ larger than for those with classical bulges and $\sim4$ larger than for those with pseudo bulges.

\item The median size (effective radius) of BCBs is $\sim2$ kpc which is $\sim$3/4th the size of classical bulges and $\sim$2/3rd the size of pseudo bulges, in both rest-bands.

\item BCBs are $\sim2.5$ times more massive than classical bulges and $\sim6$ times more massive than pseudo bulges. 

\item Star formation rate and specific star formation rate for disc galaxies with BCBs is more than a factor of $\sim2$ and $\sim1.5$ times larger than disc galaxies with other bulge types.

\item Disc galaxies with BCBs, unlike those with pseudo and classical bulges, are outliers to the star-forming main-sequence (SFR-mass plane).

\item The $B-I$ colour index for the bulge and the host disc suggests that star formation activity is more towards the host disc.

\item Only about $\sim0.2$\% of the local bulges can be classified as BCBs, suggesting their near-absence in the local Universe.

\item Unlike pseudo and classical bulges, BCB bulges show nearly no overlap with the local bulge population, suggesting a morphological transformation from intermediate redshifts to the present epoch.

\end{itemize}

Overall, BCBs are a factor of $\sim2.5$ times brighter, $\sim3/4$ times compact, $\sim2$ times dominant, $\sim2.5$ times massive than classical bulges. In comparison to pseudo bulges, they are $\sim5$ times brighter, $\sim2/3$ times compact, $\sim4$ times dominant, $\sim6$ times massive. 

Examination of the local population of bulges suggests that a very minuscule fraction ($\sim0.2$\%) can be classified as BCB. Their rarity at intermediate redshifts and near-absence in the local explains why this bulge type has not been reported so far. However, distinctness in the properties of pseudo and classical bulges has been explored in quite some detail which has revealed that possibly different physical mechanisms have been responsible for their formation and growth inside disc galaxies \citep{KormendyandKennicutt2004,Athanassoula2005,Bournaudetal2007,BrooksandChristensen2016}. 

This suggests that physical mechanisms responsible for the formation and growth of BCB like bulges are also different than those experienced by other bulge types. Some evidence towards that is obtained from their star formation activity. While spheroid dominated, massive galaxies have been observed to be passive \citep{Stratevaetal2001,Kauffmannetal2003,Brammeretal2011}, star formation rates for disc galaxies with BCBs are larger by a factor of $\sim2$ compared to other disc galaxies. The existence of a large bulge has been observed to reduce the efficiency of star formation \citep{Saintongeetal2012,Genzeletal2014,Langetal2014} by stabilizing the disc against gravitational instabilities \citep{JogandSolomon1984,Martigetal2009}. Thus, disc galaxies with BCBs, having most dominant and massive bulges, show peculiar behaviour in terms of their star formation activity.

One possibility is that disc galaxies with BCBs are descendants of massive, compact, passive elliptical galaxies observed at high redshifts, i.e., $z>1.5$, also called ``red nuggets" \citep{Daddietal2005,Trujilloetal2006,vanDokkumetal2008}. Since, these ellipticals are also reported to lack local counterparts \citep{Damjanovetal2009,Tayloretal2010}, it is likely that they evolved by forming a compact blue disc around them \citep{Oldhametal2017}. Thus, the increased star formation activity is going on in the newly acquired disc. Conforming to that we have found that star formation activity in BCB host disc galaxies is more towards the host disc which is observed to be bluer than the bulge. If this is indeed the case, it will provide evidence to the argument that later gas accretion around pre-existing spheroids should have played a significant role in disc formation \citep{Keresetal2005,Dekeletal2009a,Dekeletal2009b,Conseliceetal2013,Putman2017}.

Another possibility is that disc galaxies with BCBs are descendants of massive, compact, star forming spheroidal galaxies observed at high redshifts, also called ``blue nuggets" \citep{Barroetal2013,Barroetal2014,Barroetal2017}. In Fig.~\ref{bluenuggets}, we have plotted our galaxies on the same plane at that defined in \citet{Barroetal2013} (see their Fig.~2), where the four quadrants categorize galaxies on the basis of their compactness and star formation efficiency. While almost all disc galaxies with BCBs reside well within star-forming region, none of them resides in the compact region. This is probably because they gained a disc later. Concrete evidence requires the examination of the gradient of colour index and stellar population analysis inside these galaxies with higher resolution, which shall become possible with newer instruments capabilities.

\section*{Acknowledgments}

We acknowledge the support of CEFIPRA-IFCPAR grant through project number 5804-1. We are highly grateful to the referee for providing insightful suggestions which have improved the quality of this work. 

%%%%%%%%%%%%%%%%%%%%%%%%%%%%%%%%%%%%%%%%%%%%%%%%%%%%%%%%%%%%%%%%%%%%%%%%%%%%%%%%%%%%%%%%%%%%%%%%%%%%%%

\bsp

\label{lastpage}
\end{document}